\documentclass[floatfix,
reprint,
superscriptaddress,
 amsmath,amssymb,
 aps,
pra,
]{revtex4-2}

\usepackage{gensymb}
\usepackage{graphicx, float}% Include figure files
\usepackage{dcolumn}% Align table columns on decimal point
\usepackage{bm}% bold math
\usepackage[utf8]{inputenc}

\usepackage{graphicx}
\usepackage{physunits}
\usepackage{xcolor}
\usepackage{amsmath}
\usepackage{multirow}
\usepackage{ulem}
\usepackage{graphicx}
\usepackage{amsmath}
\usepackage{amssymb}
\usepackage{comment}
\usepackage{soul}

\begin{document}
\title{Controlled Spin-Orbit Torques in Rare-Earth Synthetic Ferrimagnets Probed by Sagnac Magneto-Optics and Harmonic Hall Measurements
%Spin Orbit Torques in Synthetic Ferrimagnets: A Comparison of Second Harmonic and Sagnac Magneto-Optic Methods
}

\author{Akilan K}
\affiliation{Institut Jean Lamour, Université de Lorraine, CNRS, F-54000 Nancy, France}

\author{Koral Aykin}
\affiliation{
 Center for Quantum Phenomena, Department of Physics, New York University, New York, NY 10003, USA}

\author{Jose-Luis Ampuero}
\affiliation{Institut Jean Lamour, Université de Lorraine, CNRS, F-54000 Nancy, France}
\author{Apurba Maji}
\affiliation{Institut Jean Lamour, Université de Lorraine, CNRS, F-54000 Nancy, France}
\author{Laurent Badie}
\affiliation{Institut Jean Lamour, Université de Lorraine, CNRS, F-54000 Nancy, France}
\author{Stephane Mangin}
\affiliation{Institut Jean Lamour, Université de Lorraine, CNRS, F-54000 Nancy, France}
\author{Sébastien Petit-Watelot}
\affiliation{Institut Jean Lamour, Université de Lorraine, CNRS, F-54000 Nancy, France}
\author{Michel Hehn}
\affiliation{Institut Jean Lamour, Université de Lorraine, CNRS, F-54000 Nancy, France}
\author{Andrew D. Kent}
\email{andy.kent@nyu.edu}
\affiliation{
 Center for Quantum Phenomena, Department of Physics, New York University, New York, NY 10003, USA}

\author{J.-Carlos Rojas-S\'anchez}
\email{juan-carlos.rojas-sanchez@univ-lorraine.fr}
\affiliation{Institut Jean Lamour, Université de Lorraine, CNRS, F-54000 Nancy, France}

\begin{abstract}
Spin-orbit torque (SOT) provides an efficient route for manipulating magnetization in spintronic devices, and its accurate quantification is essential. Here, we investigate Co/Gd-based synthetic ferrimagnets using complementary magnetotransport and magneto-optical techniques. We use a Sagnac magneto-optical interferometry method to directly quantify current-induced magnetization tilting and extract the damping-like SOT effective field. The Sagnac measurements show good agreement with harmonic Hall analysis across different Co/Gd and Gd/Pt/Co heterostructures, providing a quantitative determination of the damping-like SOT effective field that does not rely on electrical transport signatures.
By varying the Gd thickness and multilayer stacking order, we demonstrate that the damping-like torque is strongly influenced by spin transport and angular momentum conversion at rare-earth and heavy-metal interfaces. The observation of a finite torque in Gd/Pt/Co/Pt structures, where conventional Pt spin Hall contributions are expected to compensate, highlights the active role of the Gd layer in SOT generation. Furthermore, interface engineering enables perpendicular magnetic anisotropy in Gd/Pt/Co/Al heterostructures with a 2-nm-thick Co layer, allowing current-induced SOT switching at current densities of 8-10 MA/cm$^2$. These results establish rare-earth-based synthetic ferrimagnets as a versatile platform for engineering spin-orbit phenomena and optimizing low-power spintronic devices.
\end{abstract}
\maketitle

\section{Introduction}

With the rapid growth of the Internet of Things and artificial intelligence, the energy demand associated with data processing and storage is increasing exponentially and is predicted to account for up to 50$\%$ of global electricity consumption by 2030 \cite{andrae2024global,hoefflinger2020irds}. Consequently, disruptive technologies are required to reduce energy consumption in computing systems. Spintronics offers promising solutions through non-volatile magnetic random-access memory (MRAM), including spin-transfer torque (STT)-MRAM \cite{Berger1996,Slonczewski1996,Liu2010,brataas2012current} and spin-orbit torque (SOT)-MRAM \cite{Miron2010,gambardella2011current}. Advancing SOT-MRAM technology requires reducing the critical current needed to switch the magnetization while maintaining a low device resistance, thereby preserving the advantage of reduced power consumption.

Among the various classes of spintronic materials, rare-earth-transition-metal (RE-TM) ferrimagnetic alloys have attracted considerable attention because they enable both electrically driven SOT phenomena \cite{Pham2018_PhysRevAppl, Bello2022_GdFeCo} and optically induced magnetization dynamics, including current-induced self-torques \cite{cespedes2021current, Damas2022_PSS_RLL_GdFeCo}, ultrafast domain-wall motion \cite{quessab2021interplay}, and single-pulse all-optical switching (AOS) \cite{stanciu2007all,radu2011transient}. Metallic ferrimagnetic alloys such as GdFeCo, GdCo, and CoTb consist of two antiferromagnetically coupled magnetic sublattices \cite{Pham2018_PhysRevAppl, cespedes2021current, Damas2022_PSS_RLL_GdFeCo, Damas2026, Kim2022_NatureMat, Lin2023PRB}. As a result, they combine favorable characteristics of ferromagnets and antiferromagnets, while their net magnetization can be continuously tuned through composition, layer thickness, or temperature \cite{van2020deterministic,sala2022asynchronous,kim2017fast}.

Compared with amorphous RE--TM alloys, synthetic ferrimagnets (SFiMs) based on RE/TM multilayers provide greater flexibility for wafer-scale fabrication and interface engineering. Their layered structure enables independent control of the constituent layer thicknesses and interfaces, allowing the magnetic properties to be tailored over a broad range. In combination with heavy metals exhibiting strong spin-orbit coupling, such as Pt, trilayers based on Pt/Co/Gd enable independent tuning of the perpendicular magnetic anisotropy (PMA) and the interfacial Dzyaloshinskii--Moriya interaction (DMI), thereby optimizing current-driven spintronic phenomena \cite{yang2015domain,ha2016very}. More recently, Co/Gd multilayers have demonstrated robust PMA, wide-range single-shot AOS, and suppression of spin dephasing~\cite{li2023ultrafast,xie2024giant, Kunyangyuen2025ACSMatLett}. In particular, reaching the magnetic and angular momentum compensation regimes has emerged as a key strategy for enhancing both static and ultrafast magnetic properties.

Most previous studies on Co/Gd SFiMs have focused on repeated multilayer stacks composed of sub-nanometer Co and Gd layers, indicate significant Co-Gd alloying as the number of repetitions increases \cite{kim2022field}. Optimizing such heterostructures for efficient spin transport requires considering not only spin-current generation in the heavy metal but also spin generation, conversion, diffusion, and absorption within the rare-earth layer and across the corresponding interfaces. The SOT efficiency can therefore be engineered by controlling the rare-earth thickness and interface structure, potentially enabling field-free magnetization switching. Recent studies have reported strong spin absorption by spontaneously formed interfacial CoGd alloys \cite{dou2025high}, together with non-local spin torques originating from Gd/Pt interfaces \cite{dutta2024observation}. Understanding the microscopic mechanisms responsible for these torque-generation processes is therefore essential for optimizing both SOT- and AOS-based spintronic devices.

Accurate quantification of the SOT efficiency is equally important for identifying the physical origin of these effects and for comparing different material systems. Most established SOT characterization techniques have been developed for structures with in-plane magnetization and include second-harmonic Hall measurements \cite{garello2013symmetry,hayashi2014quantitative,avci2014interplay}, spin-torque ferromagnetic resonance (ST-FMR) \cite{liu2011spin,reynolds2017spin}, and magneto-optical methods based on the magneto-optical Kerr effect (MOKE) \cite{fan2014quantifying,montazeri2015magneto}. Conventional magneto-optical techniques often require corrections for the Oersted field obtained from simulations or spatially resolved measurements, increasing the complexity of the analysis. Moreover, discrepancies have been reported between the standard harmonic Hall method and alternative optical techniques, highlighting the need for independent and reliable approaches for SOT quantification.

In this work, we use a magneto-optical protocol based on Sagnac interferometry to quantify spin-orbit torques. Owing to its exceptional sensitivity, the interferometer directly probes current-induced magnetization tilting by detecting Kerr rotations with sensitivities of a few tens of nanoradians. We compare the extracted SOT efficiencies with those obtained from conventional second-harmonic Hall measurements, finding good agreement while providing a direct optical determination of the torque. We then apply the method to Co/Gd-based heterostructures composed of relatively thick magnetic layers, including Co/Gd bilayers with different stacking orders capped with Pt, as well as structures in which a Pt spacer is inserted between the Gd and Co layers and capped with either Pt or Al. Unlike the ultrathin multilayer systems commonly studied, these heterostructures allow the influence of the rare-earth layer and the interfaces on spin transport to be investigated independently. Finally, we demonstrate perpendicular magnetic anisotropy in Gd/Pt/Co/Al heterostructures containing a 2-nm-thick Co layer and observe current-induced SOT switching at current densities as low as 10 MA/cm$^2$

%In this work we measure simulatanoues longi-tudinal and transverse voltages to account for non-linear magnetoresistance effects as recently proposed by \cite{noel2025estimation,noel2025nonlinear} to obtain corrections on overestimations of the torques. This is compared with an independent measurement using a highly sensitive optical technique, the Sagnac MOKE interferometer, which is a direct optical detection of magnetic moments capable of measuring Kerr deflections of the magnitude of less than 20 nrad. The MOKE uses a single laser beam to sense the Kerr rotation and ellipticity with polarizers and analysers. The Sagnac MOKE comprises two spatially overlapping orthogonal beams that are reciprocal and travel in opposite directions along the and slow axes of a single-mode optical fiber. The changes in phase and amplitude of the left and right circularly polarized light caused by the magnetic sample is measured. This provides a direct access to any non-reciprocity effects such as time reversal symmetry breaking caused by magnetism\cite{Xia2006,Kapitulnik2009}.

\section{SOT quantification by Harmonic Hall and Sagnac MOKE }
\label{Section1}

\subsection{SOT characterization by Harmonic Hall analysis }
\label{Section1}
The second-harmonic Hall technique is one of the most widely used methods for quantifying spin-orbit torque (SOT) efficiencies~\cite{garello2013symmetry,hayashi2014quantitative,avci2014interplay,Cogulu2022}. Therefore, before introducing the magneto-optical Sagnac approach, we first characterize the SOT using the conventional second-harmonic Hall method, which serves as a reference throughout this work. 

Second-harmonic Hall measurements were performed on six-terminal Hall-bar devices. A low-frequency alternating current (433 Hz) was injected along the current channel while the longitudinal and transverse first- and second-harmonic voltages were simultaneously measured using lock-in detection. Several experimental protocols have been proposed, including magnetic-field sweeps, in-plane and out-of-plane angular measurements, and field sweeps at a fixed tilt angle~\cite{hayashi2014quantitative,avci2014interplay,yang2020characterization,xu2025alternative,Cogulu2022}. In this work, we employ the in-plane angular dependence. Because this protocol has recently been shown to exhibit nonlinear contributions arising from current-induced magnon generation and annihilation at low magnetic fields~\cite{noel2025estimation,noel2025nonlinear}, we simultaneously measure both the transverse ($V_{xy}$) and longitudinal ($V_{xx}$) voltages in a double Hall-bar geometry under a constant in-plane magnetic field while rotating its direction by an angle $\varphi_H$ (Fig.~\ref{fig2}a).

The corresponding first-harmonic, $R^{1\omega}$, and second-harmonic, $R^{2\omega}$, resistances are obtained by dividing the measured rms voltage, $V_\mathrm{rms}$, by the rms current, $I_\mathrm{rms}$, and are given by: 

\begin{equation}
\label{Eq:Rxy1w}
\begin{split}
    R^{1\omega}_{xy} =R_\mathrm{PHE}\sin(2\varphi), 
    \end{split}
\end{equation}
\begin{equation}
\label{Eq:Rxx1w}
\begin{split}
R^{1\omega}_{xx} =R^{\parallel}_\mathrm{xx}\cos^{2}\varphi+R^{\perp}_\mathrm{xx}\sin^{2}\varphi,
 \end{split}
\end{equation}
\begin{equation}
\label{Rxy2w}
\begin{split}
    R^{2\omega}_{xy} =A_{xy}\cos \varphi+B_{xy}\cos^3\varphi,
    \end{split}
\end{equation}
\begin{equation}
\label{Rxx2w}
\begin{split}
    R^{2\omega}_{xx} =A_{xx}\sin\varphi+B_{xx}\sin^3\varphi,
    \end{split}
\end{equation}
\\
where $R_\mathrm{PHE}$ is the planar Hall resistance and $R^{\parallel}_{xx}$ and $R^{\perp}_{xx}$ are longitudinal resistances when current is parallel and perpendicular to the magnetic field respectively.  

Neglecting the nonlinear magnon contribution, the coefficients $A_{xy}$ and $B_{xy}$ can be used to extract the damping-like (DL) and field-like (FL) effective fields, $B_\mathrm{DL}$ and $B_\mathrm{FL}$, together with the corresponding SOT efficiencies, $\xi^\mathrm{DL}$ and $\xi^\mathrm{FL}$. Specifically,
$R^{2\omega}_{\mathrm{DL}+\nabla{T}}=-2(A_{xy}+ \frac{B_{xy}}{2})$ and $R^{2\omega}_{\mathrm{FL}}= \frac{B_{xy}}{2}$.
By repeating the angular measurements under different applied magnetic fields, $B_\mathrm{DL}$ and $B_\mathrm{FL}$ are obtained from the linear relationships expressed in the following equations:

\begin{equation}
\begin{split}    R^{2\omega}_{\mathrm{DL},xy} =\frac{R_\mathrm{AHE}B_\mathrm{DL}}{\mu_0H_\mathrm{eff}}+R_\mathrm{thermal},
    \end{split}
        \label{Eq:RDL2omega}
\end{equation}
\begin{equation}
\begin{split}
R^{2\omega}_{\mathrm{FL},xy} =\frac{R_\mathrm{PHE}B_\mathrm{FL}}{\mu_{0}H}.
    \end{split}
    \label{FLlinear}
\end{equation}
$H_\mathrm{eff}$ is the effective magnetic field, which includes the demagnetizing and anisotropy fields in addition to the external magnetic field. The anisotropy field , $H_k$, and the anomalous Hall resistance, $R_\mathrm{AHE}$, are obtained from out-of-plane hysteresis loops, as shown in Fig.~\ref{Fig_AHE}.

Now, to account for the magnon-mediated correction, the coefficients of the second-harmonic longitudinal resistance, Eq.~\eqref{Rxx2w}, $A_{xx}+B_{xx}$, are needed. The magnon-induced resistance correction is $R_\mathrm{mag}=A_{xx}+B_{xx}$. This contribution is scaled by a factor $C_\mathrm{mag}$ to the non-linear PHE, $R^{2\omega}_\mathrm{PHE}$, defined as $R^{2\omega}_\mathrm{PHE}= C_\mathrm{mag}(R_\mathrm{mag}-R_\mathrm{off})$. The correction factor $C_\mathrm{mag}$ is obtained by iteratively minimizing the linear fits of the corrections to the damping- and field-like resistances, Eq.~\ref{Eq:RDL2omega} and Eq.~\ref{FLlinear}, as a function of $({\mu_{0}H_\mathrm{eff}})^{-1}$ and $({\mu_{0}H)}^{-1}$, respectively. The corrected resistances are therefore:

\begin{equation}
\label{Rxx2w}
\begin{split}
R^{'2\omega}_{\mathrm{DL},xy} = R^{2\omega}_{\mathrm{DL},xy}-C_\mathrm{mag}(R_\mathrm{mag}-R_\mathrm{off}),
    \\
    R^{'2\omega}_{\mathrm{FL},xy} = R^{2\omega}_{\mathrm{FL},xy}-\frac{1}{2}C_\mathrm{mag}(R_\mathrm{mag}-R_\mathrm{off}).
    \end{split}
\end{equation}

\subsection{Sample fabrication and magnetic chcracterization}

We performed transport and optical measurements on Co/Gd-based bilayer systems. 
The synthetic ferrimagnet (SFiM) samples include Co(2)/Gd(1)/Pt(2) and Co(2)/Gd(2)/Pt(2), as well as the corresponding samples with the reversed Co/Gd stacking order. Samples with a Pt insertion layer, denoted SFiM$_{\mathrm{Pt}}$, include Gd(2)/Pt(2)/Co(2)/Pt(2) and 
Gd(2)/Pt(2)/Co(2)/Al(3). The numbers in parentheses indicate the thickness of each layer in nanometers, and the substrate is on the left.
Samples were deposited by physical vapor deposition (PVD) at a base 
pressure below $2\times10^{-8}\,\mathrm{mbar}$. At the deposition rates used, the layers remain oxide-free. Deposition was performed on Si/SiO$_2$ (500 nm) substrates to prevent current shorting through the substrate. Prior to deposition, the SiO$_2$ surface was cleaned by Ar milling. The stacking order corresponds to that of the multilayer, with the first layer deposited directly on the SiO$_2$ substrate without any seed layer.
The films were patterned into Hall bars and microstrips by UV photolithography followed by ion-beam etching. Hall bars and microstrips with widths of 4, 10, and 20~$\mu$m were fabricated. 
The magnetic properties of the as-grown samples were characterized using a SQUID magnetometer. No post-deposition annealing was performed. The roughness of the buried interfaces was not measured. The only direct structural information is the HR-TEM image of Gd(2)/Pt(2)/Co(2)/Pt(2) (Appendix~A2), and interface roughness and intermixing are expected to differ between stacking orders.

\begin{figure}[h]
\centering
\includegraphics[width=0.5\textwidth]{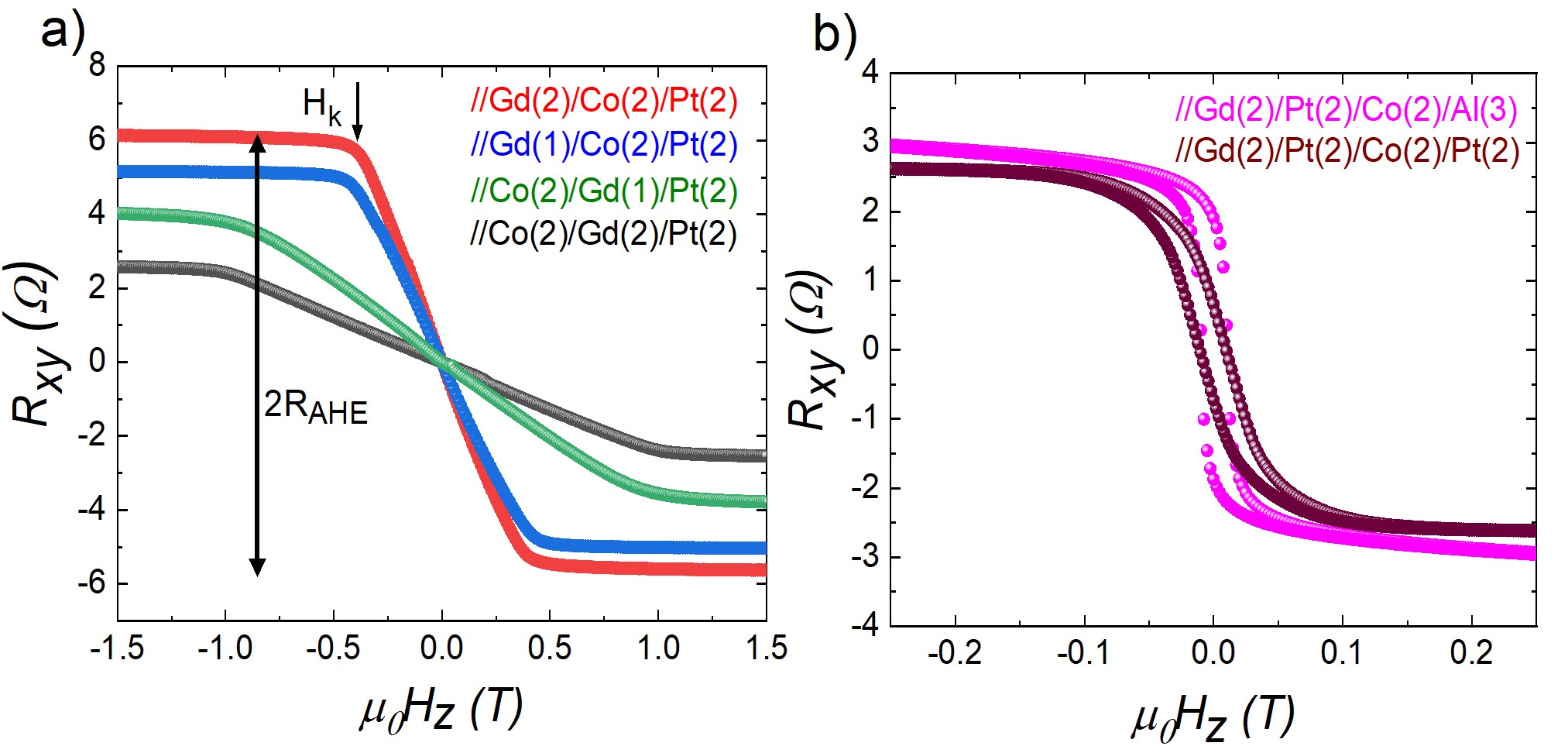}
\caption{DC measurement of anomalous Hall resistance at room temperature as a function of out-of-plane applied field in 20 ${\mu}$m width Hall bar devices in a) Co/Gd SFiM capped with Pt. b) Pt-intercalated SFiM$_{Pt}$, Gd/Pt/Co, with Al and Pt capping layers as displayed. The Gd/Pt/Co layers have a higher perpendicular anisotropy component} \label{Fig_AHE}
\end{figure}

\begin{figure}[h]
\centering
\includegraphics[width=0.5\textwidth]{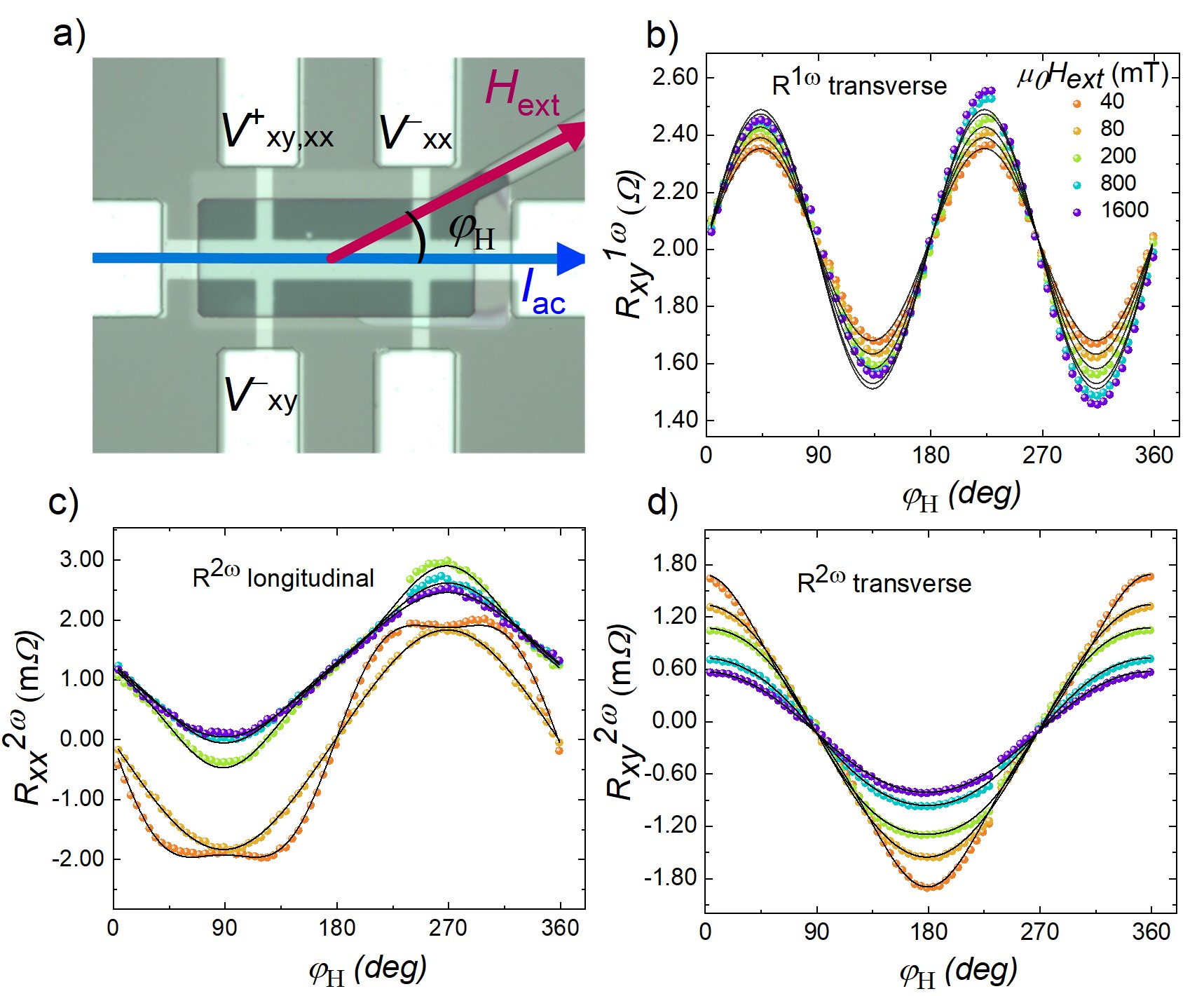}
\caption{Magnetotransport measurements in a Co(2)/Gd(1)/Pt(2) Hall bar of width 20 $\mu$m at a fixed AC current of 1 mA. Simultaneous first and second harmonics of the Hall and longitudinal resistances are shown for an in-plane angular dependence where $\phi$ is the angle between magnetization and current channel. The solid lines are fits to the equations Eqs. 1-3.}
\label{fig2}
\end{figure}

Among the Co/Gd bilayers, samples in which the Co layer is located between Gd and Pt, thereby forming a Co/Pt interface, exhibit a smaller out-of-plane saturation field, $H_k$, as shown by the red and blue curves in Fig.~\ref{Fig_AHE}(a).
This trend can be understood in terms of the balance between the demagnetizing field (shape anisotropy) and the perpendicular anisotropy constant, $K$, given by $\mu_0 H_k = \mu_0 M_s-2K/M_s$, where $M_s$ is the saturation magnetization. Since $K$ is higher for the Co/Pt interface, the resulting $H_k$ is correspondingly lower. 
An additional, complementary explanation lies in the modification of the net saturation magnetization by the adjacent layers. At the Co/Pt interface, Pt acquires a proximity-induced magnetic moment aligned parallel to the Co magnetization, thereby increasing the magnetic moment of the ferromagnetic layer. In contrast, Gd acquires a proximity-induced magnetic moment aligned antiferromagnetically to Co, which reduces the net magnetization. Consequently, the net saturation magnetization is expected to follow the approximate relation $M_{Co/Gd}<M_{Co}<M_{Co/Pt}$.
Assuming comparable values of the effective anisotropy energy, the anisotropy field scales as $\mu_0 H_k\approx2K_{eff}/M_s$, such that an increase in the effective magnetization results in a lower measured saturation field. 
This interpretation is consistent with the reduced $H_k$ observed in samples containing a Co/Pt interface.
Accordingly, all Co/Gd bilayers exhibit a well-defined in-plane magnetic easy axis, as shown in Fig.~\ref{Fig_AHE}(a), whereas the SFiMPt$_\mathrm{Pt}$ samples display a large anomalous Hall susceptibility around zero magnetic field, as shown in Fig.~\ref{Fig_AHE}(b). The enhanced susceptibility indicates that these heterostructures lie close to the spin reorientation transition, where only a modest increase in the effective perpendicular anisotropy is sufficient to stabilize an out-of-plane magnetic state. Since $t_\mathrm{Co}$ is the same in all samples, volume and surface contributions to the anisotropy cannot be separated with our data.


\subsection{Harmonic Hall measurements}
To quantify the SOT efficiency, the anomalous Hall resistance $R_\mathrm{AHE}$ and the effective magnetic field $H_\mathrm{k}$ must first be determined. These quantities are extracted from the out-of-plane Hall hysteresis loops, which provide both the anomalous Hall amplitude and the out-of-plane saturation field, $H_\mathrm{k}$, as illustrated in Fig.~\ref{Fig_AHE}(a). 

\begin{figure}[h]
\centering
\includegraphics[width=0.5\textwidth]{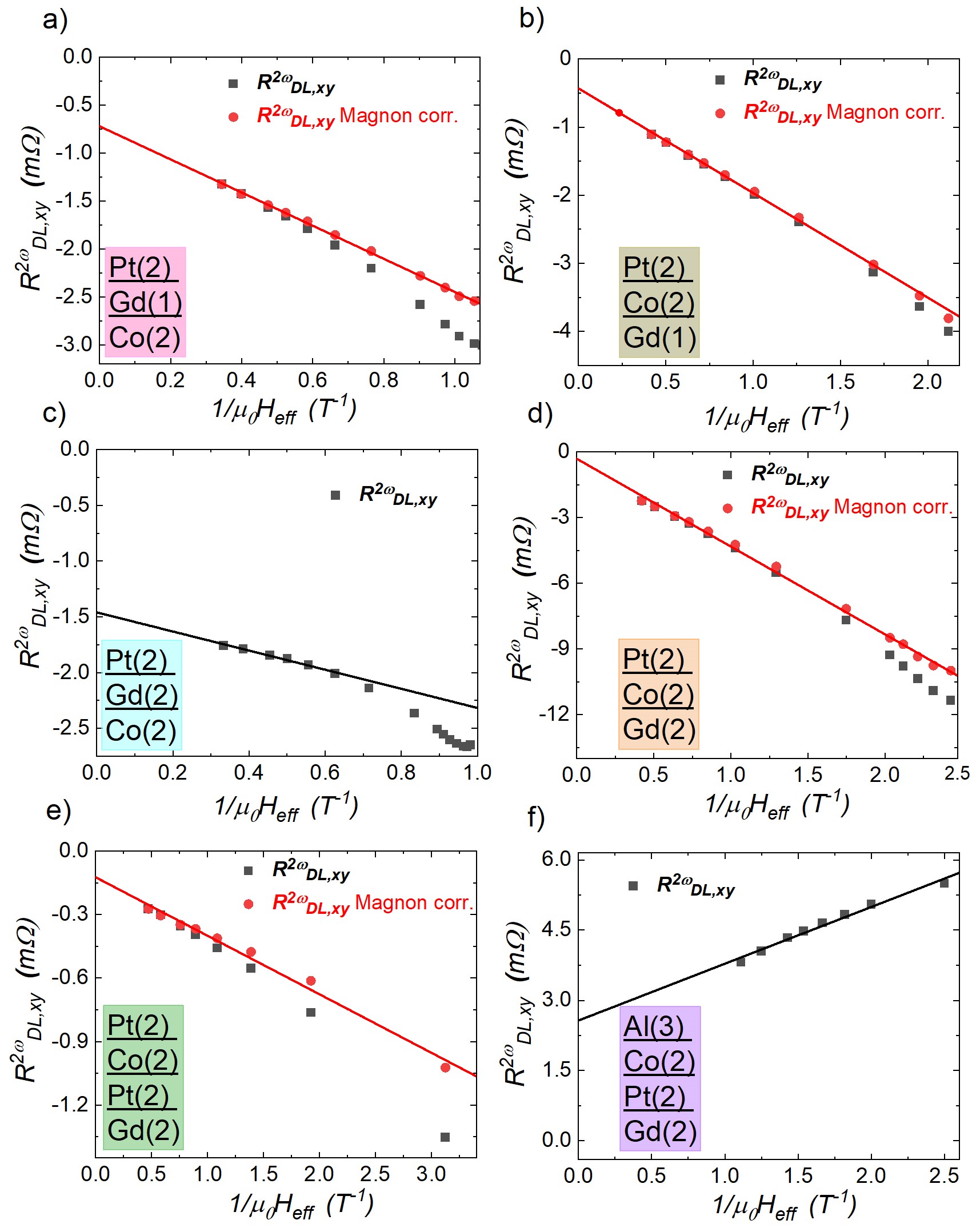}
\begin{quote}
\caption{Harmonic Hall resistance analysis of the DL resistances as a function of $({\mu_{0}H_\mathrm{eff}})^{-1}$ for an applied AC current of 1 mA peak. The correction of the non-linearity is performed in the samples a) Co(2)/Gd(1)/Pt(2), b) Gd(1)/Co(2)/Pt(2), d) Gd(2)/Co(2)/Pt(2) and e) Gd(2)/Pt(2)/Co(2)/Pt(2). The substrate is on the left.The black and red linear fits are performed for the uncorrected and corrected field points respectively. Whereas in the samples c) Co(2)/Gd(2)/Pt(2) and f) Gd(2)/Pt(2)/Co(2)/Al(3) the linear fits (black) are done without applying the magnon correction protocol \eqref{Eq:RDL2omega}.}\label{Fig_Corrections}
\end{quote}
\end{figure}

\iffalse Transport measurements were first conducted on a standard sample of Pt(5)/Py(6)/Al(3) to evaluate damping-like torques (DL) and compare these findings with the values in previous reports and results obtained from the optical Sagnac technique. Simultaneous harmonic Hall and longitudinal resistance measurements were performed on a 10 $\mu$m Hall bar. Pt is a well-studied heavy metal layer with a high positive spin Hall angle. We report a DL efficiency normalized by the current density in the Pt heavy-metal layer, following Eq.~\eqref {eff_DL_J}, as $\xi^\mathrm{DL}_{J}$= 0.0963, which aligns with values in the literature.\fi
Figure~\ref{fig2} presents the measurement protocol together with the corresponding raw data obtained from the Co(2)/Gd(1)/Pt(2) sample. As mentioned earlier, the anomalous Hall effect is measured to obtain $R_\mathrm{AHE}$ and out-of-plane saturation field $\mu_0H^\mathrm{OOP}_\mathrm{sat}$=908 mT, which provides a measure of demagnetizing fields and anisotropy fields. The angular dependence of the harmonics as shown in Fig.~\ref{fig2} is fitted with Eqs.~\ref{Eq:Rxy1w} and \ref{Eq:Rxx1w} to extract the damping-like and thermal contributions $R_{\mathrm{DL}+\nabla{T}}$. We plot $R_{\mathrm{DL}+\nabla{T}}$ against the inverse of the effective field $(\mu_{0}H_\mathrm{eff})^{-1}$ to perform linear fits. Based on standard models, the high-field regime, which shows linear behavior, is commonly used in fits, assuming the model deviates under unsaturated conditions. However, following recent reports~\cite{noel2025estimation,noel2025nonlinear}, we perform a full correction to obtain a linear behavior as shown in Fig.~\ref{Fig_Corrections}. 
The corrected data exhibit a linear dependence over the entire measured field range, allowing all field values down to 50 mT to be included in the linear regression. The correction was validated down to $\mu_0H = 50$ mT ($H \gtrsim 0.05\,H_k$). In this case, for Co(2)/Gd(1)/Pt(2) we obtain a correction factor of $C_\mathrm{mag}$= 0.65. The same $C_\mathrm{mag}$ factor is used to obtain corrected FL SOT values. Since this work focuses on comparing the Sagnac MOKE technique and transport, we report and compare only the DL SOT.
\\
\begin{figure}[h]
\centering
\includegraphics[width=0.35\textwidth]{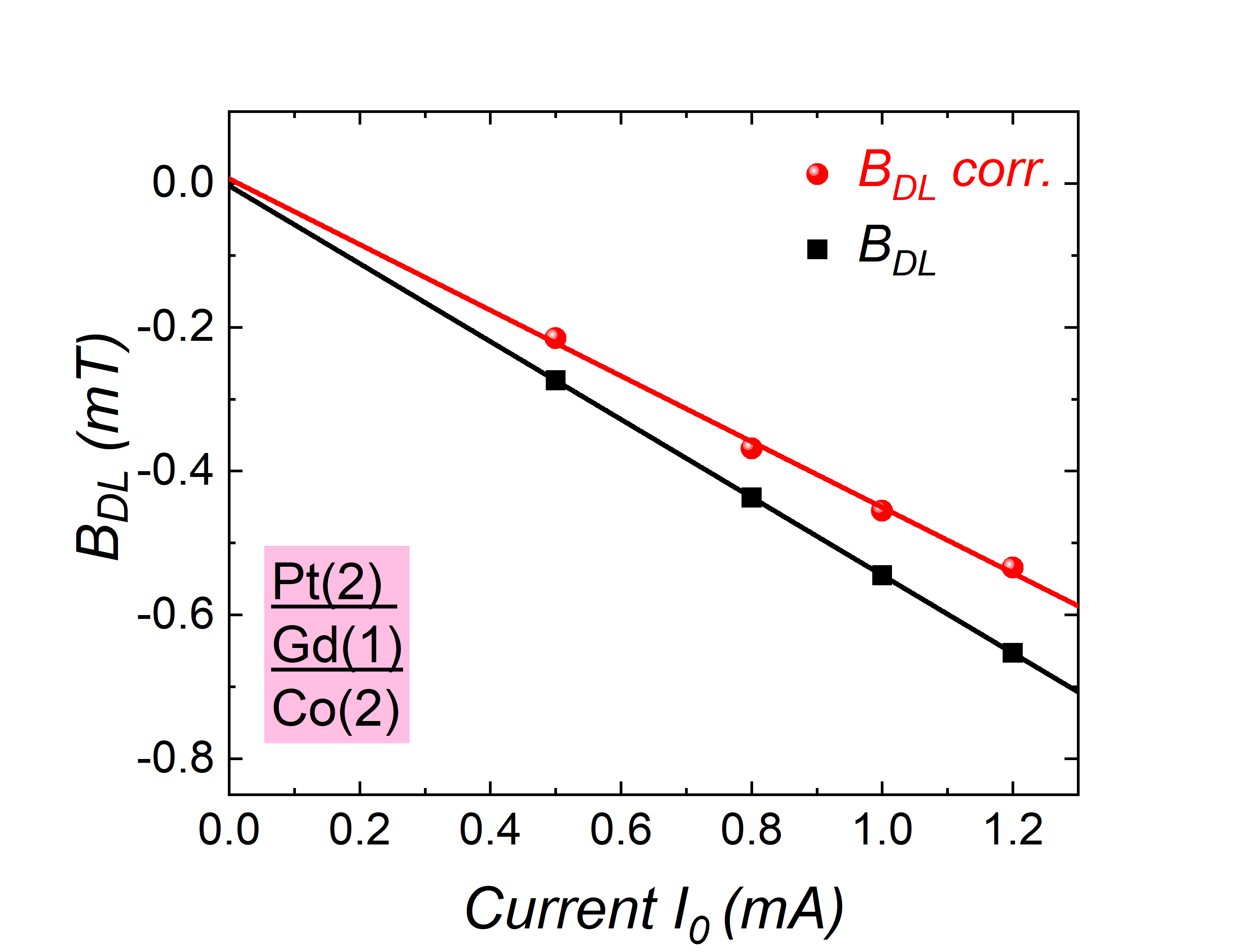}
\begin{quote}
\caption{The correction is performed for measurements at different applied currents to obtain a consistent correction factor $C_\mathrm{mag}$ for all applied currents. $B_\mathrm{DL}$ corrected (red) and uncorrected (black) for all currents is shown for the sample //Co(2)/Gd(1)/Pt(2)}\label{Fig_BDL_vs_i}
\end{quote}
\end{figure}

For the Co/Gd SFiM and SFiM$_\mathrm{Pt}$ heterostructures, we express the SOT efficiency in terms of the applied electric field [Eq.~(\ref{eff_DL_E})] rather than the total current density [Eq.~(\ref{eff_DL_J})]. This normalization avoids uncertainties associated with the current distribution among the different layers, which depends on their individual conductivities, interface scattering, and thicknesses. Consequently, the electric-field normalization provides a more robust metric for comparing samples with different multilayer architectures.
\\
 
\begin{equation}
\label{eff_DL_J}
\begin{split}
     \xi^\mathrm{DL}_{J} =\frac{2e}{\hbar}M_st_\mathrm{FM}\frac{B_\mathrm{DL}}{J_\mathrm{total}},
    \end{split}
\end{equation}
\begin{equation}
\label{eff_DL_E}
\begin{split}
     \xi^\mathrm{DL}_{E} =\frac{2e}{\hbar}M_st_\mathrm{FM}\frac{B_\mathrm{DL}w}{IR_\mathrm{sheet}}.
    \end{split}
\end{equation}

Figure~\ref{Fig_Corrections} summarizes the determination of $B_\mathrm{DL}$ after applying the magnon-correction protocol to the investigated Co/Gd-based heterostructures. Two samples constitute exceptions. For Co(2)/Gd(2)/Pt(2), the extracted $R_\mathrm{mag}$ is essentially independent of magnetic field, preventing a consistent scaling procedure; consequently, the reported value corresponds to the uncorrected analysis and should be regarded as an upper estimate. Here the correction does not capture all nonlinear contributions at $H \ll H_k$, and the uncorrected data are linear only for $H \gtrsim 0.06 H_k$.
In contrast, the Gd(2)/Pt(2)/Co(2)/Al(3) sample exhibits a large in-plane saturation field ($\mu_0H_\mathrm{sat}\approx1.1$ T), allowing the angular measurements to be performed entirely above magnetic saturation ($\mu_0 H \geq 1.5$ T). Under these conditions, magnon-induced nonlinearities are strongly suppressed, and the dependence of $R^{2\omega}_{\mathrm{DL}}$ on $(\mu_0H\mathrm{eff})^{-1}$ remains linear without requiring any correction. The value reported for this device depends on its anisotropy (Sec. III, Appendix A3). Measurements with the field-sweep protocol on a $4\mu$m-wide Hall bar are shown in Appendix A4.
The complete analysis is repeated for several applied current amplitudes to determine the final damping-like SOT efficiencies (Fig.~\ref{Fig_BDL_vs_i}). 
The Sagnac magneto-optical measurements are presented in Sec.~II\,D, and the comparison with these transport-derived values (Table~I, Fig.~8) is discussed in Sec.~IV, where good agreement between the two methods is found for most samples.

\subsection{Sagnac MOKE Results}

To independently quantify the damping-like spin-orbit torque, we used a magneto-optic Kerr effect (MOKE) based zero-area-loop Sagnac interferometer (ZALSI) operating at a wavelength of 1550 nm~\cite{Xia2006}. This high-sensitivity optical technique was pioneered by Kapitulnik for detecting time-reversal symmetry breaking in unconventional superconductors~\cite{Xia2006,Xia2006TRSB}. The interferometer measures the nonreciprocal optical phase shift associated with the polar Kerr effect while rejecting reciprocal optical effects, such as linear birefringence. Importantly, the technique can measure static Kerr rotations without requiring modulation of the sample magnetization. In its original implementation, Kapitulnik's group demonstrated a shot-noise-limited Kerr-angle sensitivity of approximately $100~\mathrm{nrad}/\sqrt{\mathrm{Hz}}$ at an optical power of only $10~\mu\mathrm{W}$~\cite{Xia2006}.

Conventional magneto-optical measurements of SOT, as reported by Montazeri \textit{et al.}~\cite{Montazeri2015}, achieved a Kerr-angle resolution of approximately $1~\mu\mathrm{rad}$ through signal modulation. More recently, Karimeddiny \textit{et al.}~\cite{karimeddiny2023sagnac} employed Sagnac interferometry with an AC-driven SOT and sideband detection to achieve a comparable Kerr-angle resolution. Thus, although the latter study employed Sagnac interferometry, its reported resolution was similar to that previously achieved using conventional MOKE and over an order of magnitude greater than the nanoradian sensitivity demonstrated by Kapitulnik's group. In contrast, our shot-noise-limited Sagnac interferometer achieves a Kerr-angle noise spectral density of approximately $300~\mathrm{nrad}/\sqrt{\mathrm{Hz}}$ at an optical power of $10~\mu\mathrm{W}$. This sensitivity enables us to resolve extremely small current-induced magnetization tilts using DC current measurements, without requiring AC modulation of the magnetic state.

An key advantage of the Sagnac MOKE technique is that it directly probes the current-induced magnetization response rather than an electrical transport signal. Consequently, unlike harmonic Hall measurements, the optical measurement is intrinsically insensitive to planar Hall contributions, anomalous Nernst effects, and recently identified magnon-mediated transport artifacts that can complicate the extraction of spin-orbit torque efficiencies~\cite{noel2025estimation}. The Sagnac measurement therefore provides a direct and complementary determination of the DL-SOT effective field.

\begin{figure}[t]
\centering
\includegraphics[width=0.45\textwidth]{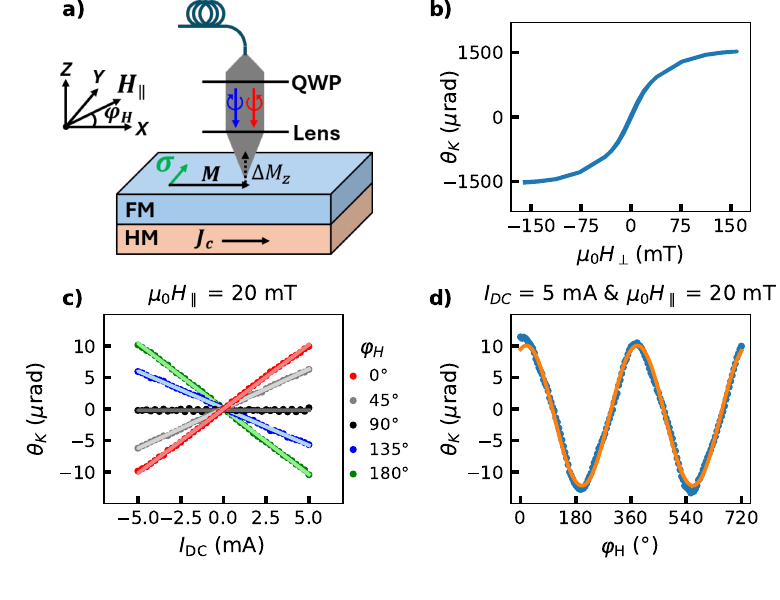}
\caption{a) Schematic of our ZALSI, configured in polar-Kerr geometry, showing a typical measurement of DL-SOT in HM/FM structures where QWP is a quarter-wave plate. b) Data for Gd(2)/Pt(2)/Co(2)/Pt(2). Kerr angle as a function of OOP field. Solid-black line is the linear fit to the data where the Kerr angle response is essentially linear. c) Kerr angle  as a function of DC current  in the presence of an IP field of 20 mT for different orientations of magnetization with respect to the current direction. d) Kerr angle as a function  of the angle ($\varphi$) between the magnetization and current direction for fixed DC of 5 mA. Solid-orange curve is the cosine fit to the data.}
\label{Fig:Sagnac1}
\end{figure}

\begin{comment}
As shown in Ref.~\cite{Xia2006}, the Kerr angle is determined from the first- and second-harmonic interferometer signals according to
\begin{equation}
\theta_K = \frac{1}{2} \tan^{-1}\left[\frac{J_2(2\phi_m)V_{1\omega}}{J_1(2\phi_m)V_{2\omega}} \right],    
\end{equation}
\noindent where $J_n$ is the $n^{th}$ Bessel function, $\phi_m$ is the electro-optic modulation depth, and $V_{1\omega}$ and $V_{2\omega}$ are the first- and second-harmonic lock-in voltages, respectively.
\end{comment}
A schematic of the measurement geometry is shown in Fig.~\ref{Fig:Sagnac1}(a). When an electric current is applied through the heavy-metal layer, a transverse spin current is generated and injected into the adjacent in-plane magnetized layer. The resulting damping-like spin-orbit torque produces a small out-of-plane tilt of the magnetization. This out-of-plane magnetization component is detected directly through the polar Kerr effect. 
Current-induced magnetization tilting has previously been employed to quantify spin-orbit torques using optical techniques~\cite{noel2025estimation,karimeddiny2023sagnac}. Here, we instead employ a quantum-shot-noise-limited Sagnac interferometer with a Kerr-angle sensitivity of approximately 300 nrad/$\sqrt{\mathrm{Hz}}$ at an optical power of 10 $\mu$W. This sensitivity enables the detection of extremely small current-induced magnetization tilts.

To quantitatively relate the measured Kerr rotation to the damping-like spin-orbit torque, we consider the equilibrium magnetization tilt produced by the SOT effective field. For a uniformly magnetized thin film strip with in-plane anisotropy, and assuming that the magnetization is aligned with the applied in-plane magnetic field ($\varphi_\mathrm{M}=\varphi_\mathrm{H}$), minimization of the magnetic free energy yields, in the small-angle limit, an equilibrium out-of-plane magnetization component of
\begin{equation}
m_\mathrm{z}=\frac{H_\mathrm{DL}\cos \varphi_\mathrm{H}}{M_\mathrm{eff}+H_\parallel},
\label{Eq:mz}
\end{equation}
where $H_\mathrm{DL}$ is the damping-like effective field, $M_\mathrm{eff}$ is the effective magnetization, and  $H_\parallel$ is the applied in-plane magnetic field. 

To quantify the damping-like effective field, we first calibrate the Kerr response by measuring the polar Kerr effect $\theta_K$ as a function of the applied perpendicular magnetic field $H_\perp$ (Fig.~\ref{Fig:Sagnac1}(b)). From this data we determine the magneto-optic susceptibility, defined as
\begin{equation}
\chi \equiv \frac{1}{\mu_0}\left.\frac{d\theta_K}{dH_\perp}\right|_{H_\perp=0},
\end{equation}
where $\theta_K$ is the Kerr rotation and $H_\perp$ is the applied out-of-plane magnetic field. 
Experimentally, $\chi$ is obtained from the slope of the Kerr rotation measured as a function of out-of-plane magnetic field in the linear regime around zero field.

Combining the expressions for the current-induced magnetization tilt and the magneto-optic susceptibility yields the relationship 
\begin{equation}
\frac{\theta_K}{\chi} = \frac{\mu_0 M_\mathrm{eff}H_\mathrm{DL}}{M_\mathrm{eff} + H_{\mathrm{\parallel}}},
\label{dl_eq}
\end{equation}
which relates the measured Kerr rotation directly to the damping-like effective field for $\varphi_H=0$, the applied current aligned with the sample magnetization. 
For most of the samples studied here, $M_\mathrm{eff} \gg H_{\parallel}$, so that the prefactor is close to unity and $\mu_0 H_{\mathrm{DL}}\simeq \theta_K/\chi$, which is therefore directly proportional to the applied current. For samples with smaller effective magnetization, the approximation $M_\mathrm{eff} \gg H_{\parallel}$ is no longer valid. In those cases, the Kerr response is measured as a function of both in-plane magnetic field and current, and Eq.~\ref{dl_eq} is fitted to extract simultaneously $M_\mathrm{eff}$ and $H_\mathrm{DL}$ (Appendix~\ref{App:Meff})

Figure~\ref{Fig:Sagnac1}(c) shows the current-induced polar Kerr rotation measured for different orientations of the applied in-plane magnetic field with respect to the current direction. As expected, the largest signal is observed for ($\varphi_H=0$), where the current and magnetization are collinear. The angular dependence of the Kerr rotation, shown in Fig.~\ref{Fig:Sagnac1}(d), follows the $\cos\varphi_H$ behavior predicted by Eq.~\ref{Eq:mz}, confirming the symmetry expected for a damping-like torque generated by the spin Hall effect.

\begin{figure}[b]
\centering
\includegraphics[width=0.4\textwidth]{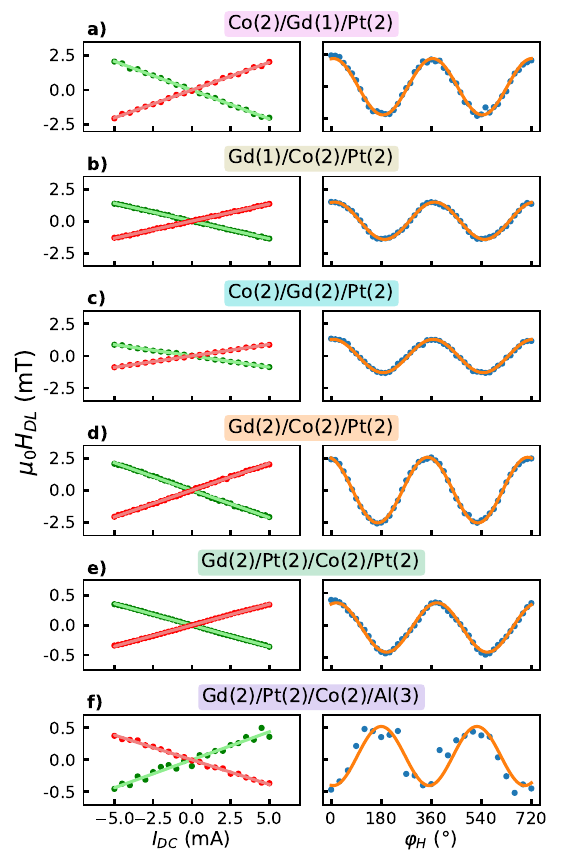}
\caption{Damping-like spin-orbit torque effective field, $\mu_0H_{\rm DL}$, determined from Sagnac MOKE. The left column shows $\mu_0H_{\rm DL}$ as a function of applied dc current, $I_{\rm DC}$, for $\mu_0H_{\parallel}=20$ mT for opposite field directions ($\phi_H=0$, red and $\phi_H=\pi$, green symbols). The right column shows the angular dependence of $\mu_0H_{\rm DL}$ as a function of the in-plane field angle, $\varphi_H$, measured at a fixed current of $I_{\rm DC}=5$ mA and $\mu_0H_{\parallel}=20$ mT. The solid orange curves are fits to a $\cos\varphi_H$ dependence expected for a damping-like torque generated by the spin Hall effect. Data are shown for (a) Co(2)/Gd(1)/Pt(2), (b) Gd(1)/Co(2)/Pt(2), (c) Co(2)/Gd(2)/Pt(2), (d) Gd(2)/Co(2)/Pt(2), (e) Gd(2)/Pt(2)/Co(2)/Pt(2), and (f) Gd(2)/Pt(2)/Co(2)/Al(3). The Gd(2)/Pt(2)/Co(2)/Al(3) sample exhibits a reversal in the sign of the damping-like effective field relative to the other heterostructures. For e), the Gd/Pt/Co/Pt sample, the small value of $M_{\mathrm{eff}}$ requires a correction to the extracted $\mu_0H_{\mathrm{DL}}$, as discussed in the Appendix; the value shown here is the uncorrected estimate obtained using the large-$M_{\mathrm{eff}}$ approximation. Table I shows the corrected value of the spin-orbit torques.
} \label{Fig:Sagnac2}
\end{figure}

Figure~\ref{Fig:Sagnac2} summarizes the damping-like effective fields extracted from the Sagnac measurements for the different Co/Gd- and Pt-based heterostructures. In every sample, the effective field increases linearly with the applied dc current, as expected for a spin-orbit torque driven by the spin Hall effect. Reversing the magnetization reverses the sign of the effective field, consistent with the symmetry of a damping-like torque. Furthermore, the angular dependence of the extracted effective field follows the expected $\cos\varphi_H$ behavior (right column), providing an independent verification that the measured Kerr rotation originates from current-induced magnetization tilting. The magnitude of the effective field depends strongly on the layer sequence and interface configuration. In particular, the largest torques are observed in the Gd(2)/Co(2)/Pt(2) and Co(2)/Gd(1)/Pt(2) heterostructures, whereas inserting a Pt spacer between Gd and Co substantially reduces the torque efficiency, highlighting the critical role of spin-current transmission across the interfaces.

Having established agreement between the Sagnac and harmonic Hall measurements, we next exploit the enhanced perpendicular magnetic anisotropy of the Gd(2)/Pt(2)/Co(2)/Al(3) heterostructure, which lies close to the spin reorientation transition, to demonstrate current-induced spin-orbit-torque switching.

\section{Current-induced SOT switching in Gd/Pt/Co/Al}
Efficient manipulation of robust perpendicular magnetization by electrical current is essential for the development of low-power magnetic memories. Among materials exhibiting high SOT efficiencies, engineering stable PMA remains a key challenge. In this work, by depositing a 2-nm-thick Gd layer directly on Si/SiO$_2$ substrates, we achieve PMA in Pt/Co-based heterostructures containing a relatively thick 2-nm Co layer.

\begin{figure}[h]
\centering
\includegraphics[width=0.5\textwidth]{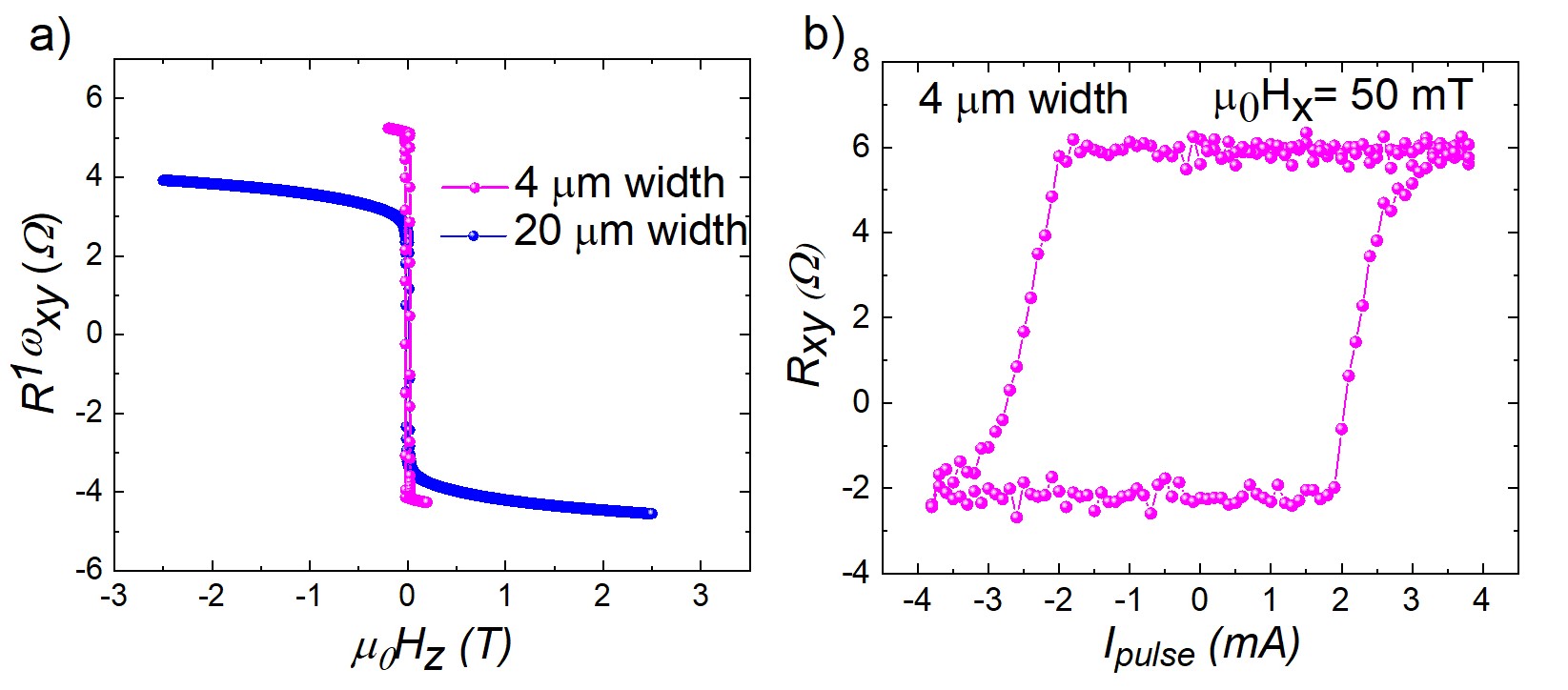}
\begin{quote}
\caption{Magnetic anisotropy and current-induced SOT switching in Gd/Pt/Co/Al: (a) Anomalous Hall resistance as a function of out-of-plane magnetic field for Hall-bar devices with 4-$\mu$m and 20-$\mu$m channel widths. The narrower device exhibits an enhanced perpendicular anisotropy component. (b) Current-induced SOT switching measured in a 4-$\mu$m-wide Hall bar under an in-plane bias field of 50 mT using 100-$\mu$s current pulses.}\label{ALcapping}
\end{quote}
\end{figure}

The as-grown Co/Gd SFiM and SFiM$_\mathrm{Pt}$ heterostructures investigated here predominantly exhibit in-plane magnetic anisotropy. However, the Gd/Pt/Co/Pt and Gd/Pt/Co/Al structures show an enhanced out-of-plane magnetic susceptibility near zero field, indicating that these systems are close to the spin reorientation transition (SRT). After device fabrication, the patterned Gd/Pt/Co/Al structures exhibit a pronounced perpendicular anisotropy component, which depends on the Hall-bar width. As shown in Fig.~\ref{ALcapping}(a), the 4-$\mu$m-wide device displays a robust PMA state, whereas the 20-$\mu$m-wide device exhibits a weaker perpendicular component.

This width-dependent anisotropy is attributed to modifications of the Co/Al interface during the lithographic processing, particularly variations in the oxidation of the Al capping layer (Appendix~A3). Previous studies have demonstrated that the magnetic anisotropy of Co/Al interfaces is strongly influenced by the oxidation state and the fraction of metallic Al remaining at the interface~\cite{krishnia2025interfacial,manchon2008x}. The effect is enhanced in narrower devices because the relative contribution of exposed and oxidized regions is larger compared with wider structures.

The induced PMA in the 4-$\mu$m-wide devices enables current-induced SOT switching, as shown in Fig.~\ref{ALcapping}(b). Despite the use of only a 2-nm-thick Pt layer, deterministic switching is achieved with critical current densities of approximately 8--10 MA/cm$^2$ under in-plane bias fields of $\mu_0H_\parallel\geq50$ mT. The resistivity of the multilayer stack is 301.5~$\mu\Omega$cm. These switching current densities are comparable to the lowest values reported for engineered Pt/Co/Al-based heterostructures~\cite{feng2020effects,miron2011perpendicular}. The relatively high resistivity observed here originates from the reduced Pt thickness and possible intermixing effects at the Gd/Pt interface, suggesting that further optimization of the layer thicknesses and interfaces could improve the electrical properties of Gd/Pt/Co-based heterostructures.

The current-induced switching remains robust under repeated current pulses and does not produce irreversible modifications of the magnetic properties, demonstrating the potential of the Gd/Pt/Co/Al platform for efficient SOT-driven magnetization manipulation.

%\newpage

\section{Discussions on Co/Gd SFiM and SFiM$_\mathrm{Pt}$ systems}

\begin{figure}[tbh]
\centering
\includegraphics[width=0.4\textwidth]{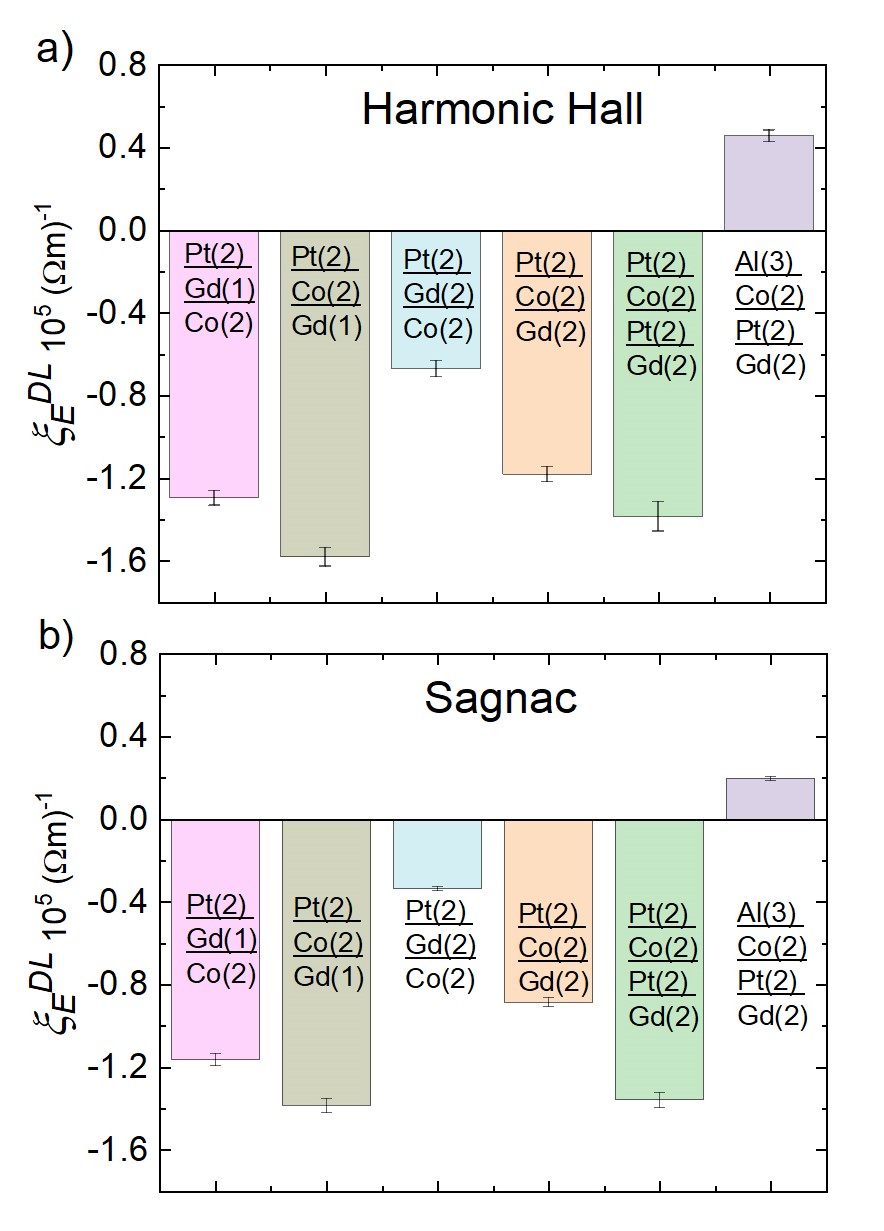}
\begin{quote}
\caption{DL SOT efficiencies per applied electric field $\xi_{\mathrm{E}}^{\mathrm{DL}}$ in the Co-Gd SFiM and SFiM$_\mathrm{Pt}$ systems based on a) harmonic Hall analysis and b) Sagnac interferometer measurements} \label{efficiencyElectricField}
\end{quote}
\end{figure}

The observation of SOT switching in Gd/Pt/Co/Al motivates a detailed comparison between the damping-like SOT efficiencies extracted from harmonic Hall and Sagnac measurements. Table \ref{table} and Figure ~\ref{efficiencyElectricField} summarizes the electric-field-normalized damping-like SOT efficiencies, $\xi_{\mathrm{E}}^{\mathrm{DL}}$, obtained from both approaches. Overall, the good agreement between the two independent techniques confirms the consistency of the extracted damping-like spin-orbit torques across all investigated heterostructures.

The good agreement between transport-based and magneto-optical measurements is particularly relevant for Co/Gd-based ferrimagnetic systems, where electrical measurements can be affected by additional nonlinear transport contributions. In contrast, the Sagnac technique directly detects the current-induced out-of-plane magnetization response and therefore provides an independent measurement of the damping-like effective field.
\\
The magnitude and sign of the extracted SOT efficiencies depend strongly on the layer sequence and interface configuration. For Co/Gd bilayers capped with Pt, increasing the Gd thickness from 1 to 2 nm reduces the damping-like SOT efficiency for both stacking orders (Table~I). In Co/Gd/Pt, where Gd lies between Pt and Co, this behavior can be attributed to the increased spin-current attenuation through the thicker Gd layer, which reduces the transmission of spin angular momentum from the Pt layer to the Co layer. In Gd/Co/Pt, Gd is not in the Pt-to-Co path, so attenuation cannot explain the reduction, which must involve the Gd/Co interface or the structural quality of the stack. For a given Gd thickness, the efficiency is also larger in Gd/Co/Pt than in Co/Gd/Pt. This is consistent with Pt growing directly on Co, but also with a different interfacial structure, since in Co/Gd/Pt the Pt grows on Gd, where intermixing is observed (Appendix~A2), and spin absorption by interfacial CoGd alloys has been reported~\cite{dou2025high}.

The Gd/Pt/Co/Pt heterostructure provides further insight into the role of the rare-earth layer in torque generation. In an ideal symmetric Pt/Co/Pt structure, the spin Hall torques generated by the two Pt layers are expected to compensate due to their opposite spin-polarization directions, resulting in a negligible net damping-like torque. However, a finite SOT is experimentally observed in the Gd/Pt/Co/Pt heterostructure, indicating that 
the two Pt layers do not fully compensate. This may reflect their structural inequivalence: HR-TEM shows a crystalline top Pt layer and a polycrystalline bottom Pt layer partially intermixed with Gd (Appendix~A2, Fig.~10), which can modify its spin Hall angle and interfacial spin transparency. It may also involve the Gd layer directly, through a bulk spin Hall effect, whose magnitude and sign are not established, through interfacial spin transmission and absorption processes at the Gd/Pt interface ~\cite{dutta2024observation} or through orbital-to-spin conversion, predicted for 4f metals~\cite{sala2022giant} and reported at Pt interfaces~\cite{Yactayo2026PRL}. In Gd/Pt/Co/Al, where only the bottom Pt layer contributes, the torque has the opposite sign and is small (Table~I), which is compatible with a reduced bottom-Pt contribution but does not identify its origin. Therefore, the measured damping-like torque cannot be described by the spin Hall effect of an ideal symmetric Pt/Co/Pt stack, but our data do not separate structural contributions from intrinsic ones of the rare-earth layer.

Disorder and a possible orbital contribution add a further ambiguity to the thickness dependence. The sheet resistance of the multilayers increases with increasing Gd thickness (Table~I), which may indicate reduced structural quality and increased disorder in thicker Gd layers, as suggested by the transmission electron microscopy analysis of Gd(2)/Pt(2)/Co(2)/Pt(2) (Appendix~A2). In clean metals with inversion and time-reversal symmetry, orbital polarization relaxes within a few atomic layers and no orbital current propagates, as shown for Ti~\cite{Yactayo2026PRL}, so the thickness dependence in clean Gd would be governed by spin transport alone. In disordered Gd, orbital transport may be enhanced. Moreover, an orbital accumulation at a Gd interface could be converted into spin in an adjacent Pt layer and carried to Co by the resulting spin current~\cite{Yactayo2026PRL}, adding to or subtracting from the Pt spin Hall torque depending on the relative signs. The observed thickness dependence of the SOT may therefore involve a combination of spin diffusion, orbital angular momentum conversion, interfacial intermixing, and disorder effects, which cannot be separated with the present data. Further studies combining structural characterization of each stack, Gd-free Pt/Co/Pt references, Gd- and Pt-thickness series, first-principles calculations of orbital and spin transport, and thickness-dependent measurements will be required to disentangle these contributions.

An important outcome of this work is that the choice of seed and interface layers provides an additional degree of freedom for engineering magnetic anisotropy. The Gd/Pt/Co/Al stacking enables perpendicular anisotropy with a relatively thick 2-nm Co layer without requiring a conventional Ta seed layer to promote (111) Pt texture. Although the measured damping-like SOT efficiency in this structure is moderate, the proximity to the spin reorientation transition reduces the effective anisotropy energy barrier and enables current-induced switching at low current densities. In addition, interfacial contributions such as Rashba-related field-like torques at the Co/Al interface may further assist the switching process~\cite{krishnia2023large}.
 
These results highlight the flexibility of RE/TM-based synthetic ferrimagnets, where magnetic anisotropy, spin transport, interface-generated torques, and orbital angular momentum conversion can be independently engineered through multilayer design. Such tunability provides additional pathways toward optimizing SOT-based devices beyond conventional heavy-metal/ferromagnet heterostructures.

\begin{widetext}
\begin{table*}[t]

\caption{Comparison of damping-like spin--orbit torque parameters obtained from harmonic Hall and Sagnac magneto-optical measurements for the investigated Co/Gd/Pt-based heterostructures. Also listed are the sheet resistance $R_{\mathrm{sheet}}$, areal magnetic moment $M_s t_{\mathrm{FM}}$, and anisotropy field $\mu_0H_k$ of each sample. The damping-like effective field per applied current $B_{\mathrm{DL}}/I$ and the electric-field-normalized damping-like efficiency $\xi_{\mathrm{E}}^{\mathrm{DL}}$ are reported for both measurement techniques along with their uncertainties. The asterisk (*) denotes harmonic Hall values that were not corrected for current-induced changes in the magnon population. Hence, the Co(2)/Gd(2)/Pt(2) value is an upper bound}
    \centering
    \begin{tabular}{|l|l|l|l|l|l|l|l|}
    \hline
        Sample & $R_\mathrm{sheet}$ & $M_{s}\times t_\mathrm{FM}$  & $\mu_0H_{k}$ &  $B_\mathrm{DL}/I$  Hall  &  $B_\mathrm{DL}/I$ Sagnac &  $\xi^{DL}_\mathrm{E}$ Hall &  $\xi^{DL}_\mathrm{E}$ Sagnac \\ \hline
        ~ & ($\Omega$) & $\mu$emu/mm$^2$ & (T) & (mT/mA)& (mT/mA) & $10^5 (\Omega $m$)^{-1}$ & $10^5 (\Omega$m$)^{-1}$  \\ \hline
       
        //Co(2)/Gd(1)/Pt(2)& 314.8 & 1.47 & \phantom{$-$}0.91 & $-$0.45 $\pm$ 0.01 &$-$0.410 $\pm$ 0.002 & $-$1.29$\pm$ 0.035&$-$1.14$\pm$ 0.03 \\ \hline
         //Gd(1)/Co(2)/Pt(2) & 203.8 & 1.72 & \phantom{$-$}0.39 &$-$0.31 $\pm$ 0.01 &$-$0.271 $\pm$ 0.001 & $-$1.58$\pm$ 0.045 &$-$1.38$\pm$ 0.035\\ \hline
        //Co(2)/Gd(2)/Pt(2) & 451.9 & 1.38 & \phantom{$-$}0.99 & $-$0.36*$\pm$ 0.02 &$-$0.177 $\pm$ 0.001 & $-$0.67*$\pm$ 0.039&$-$0.33$\pm$ 0.009 \\ \hline
        //Gd(2)/Co(2)/Pt(2) & 269.1 & 0.93 & \phantom{$-$}0.37 & $-$0.56  $\pm$ 0.01 &$-$0.421 $\pm$ 0.002 & $-$1.18 $\pm$ 0.036&$-$0.89$\pm$ 0.022\\ \hline
        //Gd(2)/Pt(2)/Co(2)/Pt(2) & 112.5 & 2.42 & \phantom{$-$}0.12 &$-$0.11  $\pm$ 0.01& $-$0.104 $\pm$ 0.001&  $-$1.38$\pm$ 0.072 & $-$1.34 $\pm$ 0.036 \\ \hline
        //Gd(2)/Pt(2)/Co(2)/Al(3)  & 497.5 & 1.98 & $-$1.1 & \phantom{$-$}0.19  $\pm$ 0.01 & \phantom{$-$}0.082 $\pm$ 0.004 & \phantom{$-$}0.46$\pm$ 0.028 & \phantom{$-$}0.27$\pm$ 0.011\\ \hline
    \end{tabular}
    \label{table}
\end{table*}
\end{widetext}

\section{Conclusions}
In this work, we investigated spin-orbit torque generation and current-induced magnetization switching in Co/Gd-based synthetic ferrimagnets using complementary magnetotransport and magneto-optical techniques. We demonstrated that Sagnac MOKE interferometry provides a highly sensitive and direct method for quantifying damping-like spin-orbit torques by measuring current-induced magnetization tilting. The Sagnac measurements are in good agreement with harmonic Hall analysis across all investigated heterostructures while providing a direct optical probe that is inherently free from many of the transport-related artifacts associated with nonlinear Hall and thermoelectric effects.

Using this methodology, we systematically studied the influence of rare-earth thickness and multilayer stacking order on the damping-like torque in Co/Gd and Gd/Pt/Co-based heterostructures. The extracted SOT efficiencies reveal a strong dependence on the location of the Gd layer and the heavy-metal interfaces, demonstrating that angular momentum transport in these systems is governed by a combination of spin generation, diffusion, absorption, and interfacial conversion processes. In particular, the observation of a finite torque in the nominally symmetric Gd/Pt/Co/Pt structure indicates that the two Pt contributions do not fully compensate. Structural asymmetry, bulk spin Hall effect in Gd and orbital-to-spin conversion are possible origins that our measurements cannot distinguish.

Furthermore, we demonstrate that interface engineering enables perpendicular magnetic anisotropy in Gd/Pt/Co/Al heterostructures containing a relatively thick 2-nm Co layer. By exploiting the proximity of this system to the spin reorientation transition, we achieve current-induced spin-orbit-torque switching with critical current densities of approximately 8-10 MA/cm$^2$ under a moderate in-plane bias field. This result highlights the importance of optimizing not only the torque magnitude but also the magnetic energy landscape for efficient SOT-driven devices.

Overall, these findings demonstrate that rare-earth-based synthetic ferrimagnets provide a versatile platform for engineering spin transport, magnetic anisotropy, and current-induced dynamics through multilayer design. The combination of direct optical torque quantification and interface-controlled magnetic engineering offers new opportunities for understanding and optimizing low-power spintronic devices based on complex spin-orbit phenomena.

\medskip
\noindent
\textbf{Acknowledgements}\\ 
This work was supported by the European Research Council (ERC) Consolidator Grant MAGNETALLIEN (Grant No. 101086807), the EU Horizon 2020 RISE project ULTIMATE-I (Ultra Thin Magneto Thermal Sensoring) (Grant No. 101007825), the French National Research Agency (ANR) through the Lorraine Université d'Excellence program (ANR-15-IDEX-04-LUE), and the W. M. Keck Foundation through its support of the research program at NYU. It was also partially supported by the ANR through the France 2030 government grants EMCOM (ANR-22-PEEL-0009), PEPR SPIN ANR-22-EXSP-0007 and ANR-22-EXSP-0009. Devices in the present study were patterned at Institut Jean Lamour's clean room facilities (MiNaLor). We thank G. Steciuk and E. Emo for their assistance with the HR-TEM characterization.

\bibliography{STT_SOT.bib}

@article{Kunyangyuen2025ACSMatLett,
author = {Kunyangyuen, Boonthum and Malinowski, Gr{\'e}gory and Lacour, Daniel and Seng, Boris and Zhang, Wei and Mangin, St{\'e}phane and Hohlfeld, Julius and Gorchon, Jon and Hehn, Michel},
title = {Controlling Single-Pulse Magnetization Switching through Angular Momentum Reservoir Engineering},
journal = {ACS Materials Letters},
volume = {7},
number = {11},
pages = {3619-3625},
year = {2025},
doi = {10.1021/acsmaterialslett.5c01060},

URL = {https://doi.org/10.1021/acsmaterialslett.5c01060},
eprint = {  https://doi.org/10.1021/acsmaterialslett.5c01060}}

@article{Lin2023PRB,
  title = {Single laser pulse induced magnetization switching in in-plane magnetized GdCo alloys},
  author = {Lin, Jun-Xiao and Hehn, Michel and Hauet, Thomas and Peng, Yi and Igarashi, Junta and Le Guen, Yann and Remy, Quentin and Gorchon, Jon and Malinowski, Gregory and Mangin, St\'ephane and Hohlfeld, Julius},
  journal = {Phys. Rev. B},
  volume = {108},
  issue = {22},
  pages = {L220403},
  numpages = {8},
  year = {2023},
  month = {Dec},
  publisher = {American Physical Society},
  doi = {10.1103/PhysRevB.108.L220403},
  url = {https://link.aps.org/doi/10.1103/PhysRevB.108.L220403}
}

@article{manchon2008x,
  title={X-ray analysis of the magnetic influence of oxygen in Pt/ Co/ AlOx trilayers},
  author={Manchon, Aur{\'e}lien and Pizzini, Stefania and Vogel, Jan and Uhlir, Vojtech and Lombard, Lucien and Ducruet, Clarisse and Auffret, St{\'e}phane and Rodmacq, Bernard and Di{\'e}ny, Bernard and Hochstrasser, Michael and others},
  journal={Journal of Applied Physics},
  volume={103},
  number={7},
  year={2008},
  publisher={AIP Publishing}
}

@article{kim2022field,
  title={Field-free switching of magnetization by tilting the perpendicular magnetic anisotropy of Gd/Co multilayers},
  author={Kim, Hyun-Joong and Moon, Kyoung-Woong and Tran, Bao Xuan and Yoon, Seongsoo and Kim, Changsoo and Yang, Seungmo and Ha, Jae-Hyun and An, Kyongmo and Ju, Tae-Seong and Hong, Jung-Il and others},
  journal={Advanced Functional Materials},
  volume={32},
  number={26},
  pages={2112561},
  year={2022},
  publisher={Wiley Online Library}
}

@Article{Kim2022_NatureMat,
author={Kim, Se Kwon
and Beach, Geoffrey S. D.
and Lee, Kyung-Jin
and Ono, Teruo
and Rasing, Theo
and Yang, Hyunsoo},
title={Ferrimagnetic spintronics},
journal={Nature Materials},
year={2022},
month={Jan},
day={01},
volume={21},
number={1},
pages={24-34},
issn={1476-4660},
doi={10.1038/s41563-021-01139-4},
url={https://doi.org/10.1038/s41563-021-01139-4}
}

@article{Bello2022_GdFeCo,
    author = {Bello, Jean-Loïs and Quessab, Yassine and Xu, Jun-Wen and Vergès, Maxime and Damas, Héloïse and Petit-Watelot, Sébastien and Rojas Sánchez, Juan-Carlos and Hehn, Michel and Kent, Andrew D. and Mangin, Stéphane},
    title = {Field-free current-induced magnetization switching in GdFeCo: A competition between spin–orbit torques and Oersted fields},
    journal = {Journal of Applied Physics},
    volume = {132},
    number = {8},
    pages = {083903},
    year = {2022},
    month = {08},
    issn = {0021-8979},
    doi = {10.1063/5.0091944},
    url = {https://doi.org/10.1063/5.0091944},
    }

@article{Damas2022_PSS_RLL_GdFeCo,
author = {Damas, Héloïse and Anadon, Alberto and Céspedes-Berrocal, David and Alegre-Saenz, Junior and Bello, Jean-Loïs and Arriola-Córdova, Aldo and Migot, Sylvie and Ghanbaja, Jaafar and Copie, Olivier and Hehn, Michel and Cros, Vincent and Petit-Watelot, Sébastien and Rojas-Sánchez, Juan-Carlos},
title = {Ferrimagnet GdFeCo Characterization for Spin-Orbitronics: Large Field-Like and Damping-Like Torques},
journal = {physica status solidi (RRL) – Rapid Research Letters},
volume = {16},
number = {6},
pages = {2200035},
doi = {https://doi.org/10.1002/pssr.202200035},
year = {2022}
}

@article{Pham2018_PhysRevAppl,
  title = {Thermal Contribution to the Spin-Orbit Torque in Metallic-Ferrimagnetic Systems},
  author = {Pham, Thai Ha and Je, S.-G. and Vallobra, P. and Fache, T. and Lacour, D. and Malinowski, G. and Cyrille, M. C. and Gaudin, G. and Boulle, O. and Hehn, M. and Rojas-S\'anchez, J.-C. and Mangin, S.},
  journal = {Phys. Rev. Appl.},
  volume = {9},
  issue = {6},
  pages = {064032},
  numpages = {9},
  year = {2018},
  month = {Jun},
  publisher = {American Physical Society},
  doi = {10.1103/PhysRevApplied.9.064032},
  url = {https://link.aps.org/doi/10.1103/PhysRevApplied.9.064032}
}

@article{Damas2026,
  title = {Spin-current symmetries generated by the ferrimagnet $\mathrm{Gd}$-$\mathrm{Fe}$-$\mathrm{Co}$ across its magnetization-compensation temperature},
  author = {Damas, H\'elo\"{\i}se and Hehn, Michel and Rojas-S\'anchez, Juan-Carlos and Petit-Watelot, S\'ebastien},
  journal = {Phys. Rev. Appl.},
  volume = {25},
  issue = {6},
  pages = {064037},
  numpages = {10},
  year = {2026},
  month = {Jun},
  publisher = {American Physical Society},
  doi = {10.1103/j438-v6q9},
  url = {https://link.aps.org/doi/10.1103/j438-v6q9}
}

@article{krishnia2023large,
  title={Large interfacial Rashba interaction generating strong spin--orbit torques in atomically thin metallic heterostructures},
  author={Krishnia, Sachin and Sassi, Yanis and Ajejas, Fernando and Sebe, Nicolas and Reyren, Nicolas and Collin, Sophie and Denneulin, Thibaud and Kov{\'a}cs, Andr{\'a}s and Dunin-Borkowski, Rafal E and Fert, Albert and others},
  journal={Nano Letters},
  volume={23},
  number={15},
  pages={6785--6791},
  year={2023},
  publisher={ACS Publications}
}

@article{miron2011perpendicular,
  title={Perpendicular switching of a single ferromagnetic layer induced by in-plane current injection},
  author={Miron, Ioan Mihai and Garello, Kevin and Gaudin, Gilles and Zermatten, Pierre-Jean and Costache, Marius V and Auffret, St{\'e}phane and Bandiera, S{\'e}bastien and Rodmacq, Bernard and Schuhl, Alain and Gambardella, Pietro},
  journal={Nature},
  volume={476},
  number={7359},
  pages={189--193},
  year={2011},
  publisher={Nature Publishing Group UK London}
}

@article{feng2020effects,
  title={Effects of oxidation of top and bottom interfaces on the electric, magnetic, and spin-Orbit torque properties of Pt/Co/Al O x trilayers},
  author={Feng, Junxiao and Grimaldi, Eva and Avci, Can Onur and Baumgartner, Manuel and Cossu, Giovanni and Rossi, Antonella and Gambardella, Pietro},
  journal={Physical Review Applied},
  volume={13},
  number={4},
  pages={044029},
  year={2020},
  publisher={APS}
}

@article{xu2025alternative,
  title={Alternative harmonic detection approach for quantitative determination of spin and orbital torques},
  author={Xu, Y and Bony, B and Krishnia, S and Torr{\~a}o Victor, R and Collin, S and Fert, A and George, J-M and Cros, V and Jaffr{\`e}s, Henri},
  journal={Applied Physics Letters},
  volume={126},
  number={20},
  year={2025},
  publisher={AIP Publishing}
}

@article{brataas2012current,
  title={Current-induced torques in magnetic materials},
  author={Brataas, Arne and Kent, Andrew D and Ohno, Hideo},
  journal={Nature materials},
  volume={11},
  number={5},
  pages={372--381},
  year={2012},
  publisher={Nature Publishing Group UK London}
}

@article{krishnia2025interfacial,
  title={Interfacial spin-orbitronic effects controlled by different oxidation levels at the Co/Al interface},
  author={Krishnia, Sachin and Voj{\'a}{\v{c}}ek, Libor and Gomes, Tristan da C{\^a}mara Santa Clara and Sebe, Nicolas and Ibrahim, Fatima and Li, Jing and Vicente-Arche, Luis Moreno and Collin, Sophie and Denneulin, Thibaud and Dunin-Borkowski, Rafal E and others},
  journal={Physical Review Applied},
  volume={24},
  number={2},
  pages={024055},
  year={2025},
  publisher={APS}
}

@article{yang2020characterization,
  title={Characterization of spin-orbit torque and thermoelectric effects via coherent magnetization rotation},
  author={Yang, Huanglin and Chen, Huanjian and Tang, Meng and Hu, Shuai and Qiu, Xuepeng},
  journal={Physical Review B},
  volume={102},
  number={2},
  pages={024427},
  year={2020},
  publisher={APS}
}

@article{reynolds2017spin,
  title={Spin Hall torques generated by rare-earth thin films},
  author={Reynolds, Neal and Jadaun, Priyamvada and Heron, John T and Jermain, Colin L and Gibbons, Jonathan and Collette, Robyn and Buhrman, RA and Schlom, DG and Ralph, DC},
  journal={Physical Review B},
  volume={95},
  number={6},
  pages={064412},
  year={2017},
  publisher={APS}
}

@article{liu2011spin,
  title={Spin-torque ferromagnetic resonance induced by the spin Hall effect},
  author={Liu, Luqiao and Moriyama, Takahiro and Ralph, D\_C and Buhrman, R\_A},
  journal={Physical review letters},
  volume={106},
  number={3},
  pages={036601},
  year={2011},
  publisher={APS}
}

@article{radu2011transient,
  title={Transient ferromagnetic-like state mediating ultrafast reversal of antiferromagnetically coupled spins},
  author={Radu, I and Vahaplar, K and Stamm, C and Kachel, T and Pontius, N and D{\"u}rr, HA and Ostler, TA and Barker, J and Evans, RFL and Chantrell, RW and others},
  journal={Nature},
  volume={472},
  number={7342},
  pages={205--208},
  year={2011},
  publisher={Nature Publishing Group UK London}
}

@article{stanciu2007all,
  title={All-optical magnetic recording with circularly polarized light},
  author={Stanciu, Claudiu D and Hansteen, Fredrik and Kimel, Alexey V and Kirilyuk, Andrei and Tsukamoto, Arata and Itoh, Achioshi and Rasing, Th},
  journal={Physical review letters},
  volume={99},
  number={4},
  pages={047601},
  year={2007},
  publisher={APS}
}

@article{gambardella2011current,
  title={Current-induced spin--orbit torques},
  author={Gambardella, Pietro and Miron, Ioan Mihai},
  journal={Philosophical Transactions of the Royal Society A: Mathematical, Physical and Engineering Sciences},
  volume={369},
  number={1948},
  pages={3175--3197},
  year={2011},
  publisher={The Royal Society}
}

@misc{andrae2024global,
  title={On global electricity usage of communication technology: trends to 2030. Challenges 2015, 6, 117-157},
  author={Andrae, ASG and Edler, T},
  year={2015}
}

@article{hoefflinger2020irds,
  title={Irds—international roadmap for devices and systems, rebooting computing, s3s},
  author={Hoefflinger, Bernd},
  journal={NANO-CHIPS 2030: On-Chip AI for an Efficient Data-Driven World},
  pages={9--17},
  year={2020},
  publisher={Springer}
}

@article{yang2015domain,
  title={Domain-wall velocities of up to 750 m s- 1 driven by exchange-coupling torque in synthetic antiferromagnets},
  author={Yang, See-Hun and Ryu, Kwang-Su and Parkin, Stuart},
  journal={Nature nanotechnology},
  volume={10},
  number={3},
  pages={221--226},
  year={2015},
  publisher={Nature Publishing Group UK London}
}

@article{ha2016very,
  title={Very large domain wall velocities in Pt/Co/GdOx and Pt/Co/Gd trilayers with Dzyaloshinskii-Moriya interaction},
  author={Ha Pham, Thai and Vogel, J and Sampaio, J and Va{\v{n}}atka, M and Rojas-S{\'a}nchez, J-C and Bonfim, M and Chaves, DS and Choueikani, F and Ohresser, P and Otero, E and others},
  journal={Europhysics Letters},
  volume={113},
  number={6},
  pages={67001},
  year={2016},
  publisher={EDP Sciences, IOP Publishing and Societ{\`a} Italiana di Fisica}
}

@article{van2020deterministic,
  title={Deterministic all-optical magnetization writing facilitated by non-local transfer of spin angular momentum},
  author={van Hees, Youri LW and van de Meugheuvel, Paul and Koopmans, Bert and Lavrijsen, Reinoud},
  journal={Nature communications},
  volume={11},
  number={1},
  pages={3835},
  year={2020},
  publisher={Nature Publishing Group UK London}
}

@article{sala2022asynchronous,
  title={Asynchronous current-induced switching of rare-earth and transition-metal sublattices in ferrimagnetic alloys},
  author={Sala, Giacomo and Lambert, Charles-Henri and Finizio, Simone and Raposo, Victor and Krizakova, Viola and Krishnaswamy, Gunasheel and Weigand, Markus and Raabe, J{\"o}rg and Rossell, Marta D and Martinez, Eduardo and others},
  journal={Nature Materials},
  volume={21},
  number={6},
  pages={640--646},
  year={2022},
  publisher={Nature Publishing Group UK London}
}

@article{kim2017fast,
  title={Fast domain wall motion in the vicinity of the angular momentum compensation temperature of ferrimagnets},
  author={Kim, Kab-Jin and Kim, Se Kwon and Hirata, Yuushou and Oh, Se-Hyeok and Tono, Takayuki and Kim, Duck-Ho and Okuno, Takaya and Ham, Woo Seung and Kim, Sanghoon and Go, Gyoungchoon and others},
  journal={Nature materials},
  volume={16},
  number={12},
  pages={1187--1192},
  year={2017},
  publisher={Nature Publishing Group UK London}
}

@article{cespedes2021current,
  title={Current-induced spin torques on single GdFeCo magnetic layers},
  author={C{\'e}spedes-Berrocal, David and Damas, Helo{\"\i}se and Petit-Watelot, S{\'e}bastien and Maccariello, Davide and Tang, Ping and Arriola-C{\'o}rdova, Aldo and Vallobra, Pierre and Xu, Yong and Bello, Jean-Lo{\"\i}s and Martin, Elodie and others},
  journal={Advanced Materials},
  volume={33},
  number={12},
  pages={2007047},
  year={2021},
  publisher={Wiley Online Library}
}

@article{quessab2021interplay,
  title={Interplay between Spin-Orbit Torques and Dzyaloshinskii-Moriya Interactions in Ferrimagnetic Amorphous Alloys},
  author={Quessab, Yassine and Xu, Jun-Wen and Morshed, Md Golam and Ghosh, Avik W and Kent, Andrew D},
  journal={Advanced Science},
  volume={8},
  number={18},
  pages={2100481},
  year={2021},
  publisher={Wiley Online Library}
}

@article{Xia2006,
    author = {Xia, Jing and Beyersdorf, Peter T. and Fejer, M. M. and Kapitulnik, Aharon},
    title = {Modified Sagnac interferometer for high-sensitivity magneto-optic measurements at cryogenic temperatures},
    journal = {Applied Physics Letters},
    volume = {89},
    number = {6},
    pages = {062508},
    year = {2006},
    month = {08},
    issn = {0003-6951},
    doi = {10.1063/1.2336620},
    url = {https://doi.org/10.1063/1.2336620},
    }

@article{Yactayo2026PRL,
  title = {Orbitronics: orbit-spin interconversions for long-range orbital propagation enabled by spin current},
  author = {Yactayo, Melissa and Pezo, A. and Ampuero, J. L. and Tian, M. and Badie, L. and Quispe-Marcatoma, J. and Landauro, C. V. and Xu, Y. and Petit-Watelot, Sébastien and Hehn, Michel and Fert, A. and Rojas-Sánchez, J. -C.},
  journal = {Phys. Rev. Lett.},
  pages = {in press},
  year = {2026},
  month = {Sep},
  publisher = {American Physical Society},
  doi = {10.1103/621t-vx16},
  url = {https://link.aps.org/doi/10.1103/621t-vx16}
}

@article{montazeri2015magneto,
  title={Magneto-optical investigation of spin--orbit torques in metallic and insulating magnetic heterostructures},
  author={Montazeri, Mohammad and Upadhyaya, Pramey and Onbasli, Mehmet C and Yu, Guoqiang and Wong, Kin L and Lang, Murong and Fan, Yabin and Li, Xiang and Khalili Amiri, Pedram and Schwartz, Robert N and others},
  journal={Nature communications},
  volume={6},
  number={1},
  pages={8958},
  year={2015},
  publisher={Nature Publishing Group UK London}
}

@article{fan2014quantifying,
  title={Quantifying interface and bulk contributions to spin--orbit torque in magnetic bilayers},
  author={Fan, Xin and Celik, Halise and Wu, Jun and Ni, Chaoying and Lee, Kyung-Jin and Lorenz, Virginia O and Xiao, John Q},
  journal={Nature Communications},
  volume={5},
  number={1},
  pages={3042},
  year={2014},
  publisher={Nature Publishing Group UK London}
}

@article{li2023ultrafast,
  title={Ultrafast Racetrack Based on Compensated Co/Gd-Based Synthetic Ferrimagnet with All-Optical Switching},
  author={Li, Pingzhi and Kools, Thomas J and Koopmans, Bert and Lavrijsen, Reinoud},
  journal={Advanced Electronic Materials},
  volume={9},
  number={1},
  pages={2200613},
  year={2023},
  publisher={Wiley Online Library}
}

@article{xie2024giant,
  title={Giant Spin--Orbit Torque in Antiferromagnetic-Coupled Pt/[Co/Gd] N Multilayers with Suppressed Spin Dephasing and Robust Thermal Stability},
  author={Xie, Zhicheng and Yang, Yumin and Chen, Bingyu and Zhao, Zhiyuan and Qin, Hongrui and Sun, Hongli and Lei, Na and Zhao, Jianhua and Wei, Dahai},
  journal={ACS Applied Materials \& Interfaces},
  volume={16},
  number={21},
  pages={27944--27951},
  year={2024},
  publisher={ACS Publications}
}

@article{dou2025high,
  title={High spin--orbit torque efficiency induced by engineering spin absorption for fully electric-driven magnetization switching},
  author={Dou, Pengwei and Zhang, Jingyan and Zhu, Tao and Kang, Peng and Deng, Xiao and Wang, Yuanbo and Qiu, Quangao and Feng, Liangyu and Hu, Jinhu and Shen, Jianxin and others},
  journal={Materials Horizons},
  volume={12},
  number={8},
  pages={2554--2563},
  year={2025},
  publisher={Royal Society of Chemistry}
}

@article{dutta2024observation,
  title={Observation of interface-induced nonlocal spin-torques from Gd/Pt interface},
  author={Dutta, Sutapa and Tulapurkar, Ashwin A and Bose, Arnab},
  journal={Applied Physics Letters},
  volume={125},
  number={22},
  year={2024},
  publisher={AIP Publishing}
}

@article{karimeddiny2023sagnac,
  title={Sagnac interferometry for high-sensitivity optical measurements of spin-orbit torque},
  author={Karimeddiny, Saba and Cham, Thow Min Jerald and Smedley, Orion and Ralph, Daniel C and Luo, Yunqiu Kelly},
  journal={Science Advances},
  volume={9},
  number={36},
  pages={eadi9039},
  year={2023},
  publisher={American Association for the Advancement of Science}
}

@article{hayashi2014quantitative,
  title={Quantitative characterization of the spin-orbit torque using harmonic Hall voltage measurements},
  author={Hayashi, Masamitsu and Kim, Junyeon and Yamanouchi, Michihiko and Ohno, Hideo},
  journal={Physical Review B},
  volume={89},
  number={14},
  pages={144425},
  year={2014},
  publisher={APS}
}

@article{avci2014interplay,
  title={Interplay of spin-orbit torque and thermoelectric effects in ferromagnet/normal-metal bilayers},
  author={Avci, Can Onur and Garello, Kevin and Gabureac, Mihai and Ghosh, Abhijit and Fuhrer, Andreas and Alvarado, Santos F and Gambardella, Pietro},
  journal={Physical Review B},
  volume={90},
  number={22},
  pages={224427},
  year={2014},
  publisher={APS}
}

@article{garello2013symmetry,
  title={Symmetry and magnitude of spin--orbit torques in ferromagnetic heterostructures},
  author={Garello, Kevin and Miron, Ioan Mihai and Avci, Can Onur and Freimuth, Frank and Mokrousov, Yuriy and Bl{\"u}gel, Stefan and Auffret, St{\'e}phane and Boulle, Olivier and Gaudin, Gilles and Gambardella, Pietro},
  journal={Nature nanotechnology},
  volume={8},
  number={8},
  pages={587--593},
  year={2013},
  publisher={Nature Publishing Group UK London}
}

@article{sala2022giant,
  title={Giant orbital Hall effect and orbital-to-spin conversion in 3 d, 5 d, and 4 f metallic heterostructures},
  author={Sala, Giacomo and Gambardella, Pietro},
  journal={Physical Review Research},
  volume={4},
  number={3},
  pages={033037},
  year={2022},
  publisher={APS}
}

@article{noel2025estimation,
  title={Estimation of spin-orbit torques in the presence of current-induced magnon creation and annihilation},
  author={Noel, Paul and Karad{\v{z}}a, Emir and Schlitz, Richard and Welter, Pol and Lambert, Charles-Henri and Nessi, Luca and Binda, Federico and Degen, Christian L and Gambardella, Pietro},
  journal={Physical Review B},
  volume={111},
  number={14},
  pages={144409},
  year={2025},
  publisher={APS}
}

@article{noel2025nonlinear,
  title={Nonlinear longitudinal and transverse magnetoresistances due to current-induced magnon creation-annihilation processes},
  author={No{\"e}l, Paul and Schlitz, Richard and Karad{\v{z}}a, Emir and Lambert, Charles-Henri and Nessi, Luca and Binda, Federico and Gambardella, Pietro},
  journal={Physical Review Letters},
  volume={134},
  number={14},
  pages={146701},
  year={2025},
  publisher={APS}
}

@article{Slonczewski1996,
	author={J. C. Slonczewski},
	title={{Current-driven excitation of magnetic multilayers}},
	journal={J. Magn. and Magn. Mat.},
	volume= 159,
	year= 1996,
	pages={L1}
}

@article{Berger1996,
	author={L. Berger},
	title={{Emission of spin waves by a magnetic multilayer traversed by a current}},
	journal={Phys. Rev. B},
	volume= 54,
	year= 1996,
	pages= 9353 
}

@article{Miron2010,
Author = {Mihai Miron, Ioan and Gaudin, Gilles and Auffret, Stephane and Rodmacq,
   Bernard and Schuhl, Alain and Pizzini, Stefania and Vogel, Jan and
   Gambardella, Pietro},
Title = {{Current-driven spin torque induced by the Rashba effect in a ferromagnetic metal layer}},
Journal = {Nature Materials},
Year = {2010},
Volume = {9},
Number = {3},
Pages = {230},
Month = {MAR},
DOI = {10.1038/NMAT2613},
ISSN = {1476-1122},
Unique-ID = {ISI:000274700900018}
}

@article{Liu2010,
author = {Liu,H.  and Bedau,D.  and Backes,D.  and Katine,J. A.  and Langer,J.  and Kent,A. D. },
title = {Ultrafast switching in magnetic tunnel junction based orthogonal spin transfer devices},
journal = {Applied Physics Letters},
volume = {97},
number = {24},
pages = {242510},
year = {2010},
doi = {10.1063/1.3527962},
URL = {https://doi.org/10.1063/1.3527962}
}

@article{Cogulu2022,
  title = {Quantifying Spin-Orbit Torques in Antiferromagnet--Heavy-Metal Heterostructures},
  author = {Cogulu, Egecan and Zhang, Hantao and Statuto, Nahuel N. and Cheng, Yang and Yang, Fengyuan and Cheng, Ran and Kent, Andrew D.},
  journal = {Phys. Rev. Lett.},
  volume = {128},
  issue = {24},
  pages = {247204},
  numpages = {6},
  year = {2022},
  month = {Jun},
  publisher = {American Physical Society},
  doi = {10.1103/PhysRevLett.128.247204},
  url = {https://link.aps.org/doi/10.1103/PhysRevLett.128.247204}
}

@article{Xia2006TRSB,
  author = {Xia, Jing and Maeno, Yoshiteru and
            Beyersdorf, Peter T. and Fejer, M. M.
            and Kapitulnik, Aharon},
  title = {High Resolution Polar Kerr Effect Measurements
           of {Sr$_2$RuO$_4$}: Evidence for Broken
           Time-Reversal Symmetry in the Superconducting State},
  journal = {Physical Review Letters},
  volume = {97},
  pages = {167002},
  year = {2006},
  doi = {10.1103/PhysRevLett.97.167002}
}

@article{Montazeri2015, 
year = {2015}, 
title = {{Magneto-optical investigation of spin-orbit torques in metallic and insulating magnetic heterostructures}}, 
author = {Montazeri, M. and Upadhyaya, P. and Onbasli, M. C. and Yu, G. and Wong, K. L. and Lang, M. and Fan, Y. and Li, X. and Amiri, P. Khalili and Schwartz, R. N. and Ross, C. A. and Wang, K. L.}, 
journal = {Nature Communications}, 
doi = {10.1038/ncomms9958}, 
pages = {8958}, 
volume = {6}
}

\appendix

\clearpage 

\section{Appendix}
\subsection{Determination of $M_\mathrm{eff}$}
\label{App:Meff}
The analysis presented in the main text assumes that
$M_{\mathrm{eff}} \gg H_{\parallel}$, so that Eq.~\ref{dl_eq}
reduces to $\mu_0 H_{\mathrm{DL}} \simeq \theta_K/\chi$.
This approximation is well justified for most of the samples
investigated here because their effective magnetizations are much
larger than the applied in-plane field ($\mu_0 H_\parallel=20$ mT). However, for samples with
a relatively small $M_{\mathrm{eff}}$, the full expression,
Eq.~\ref{dl_eq}, must be used.

\begin{figure}[H]
\centering
\includegraphics[width=0.5\textwidth]{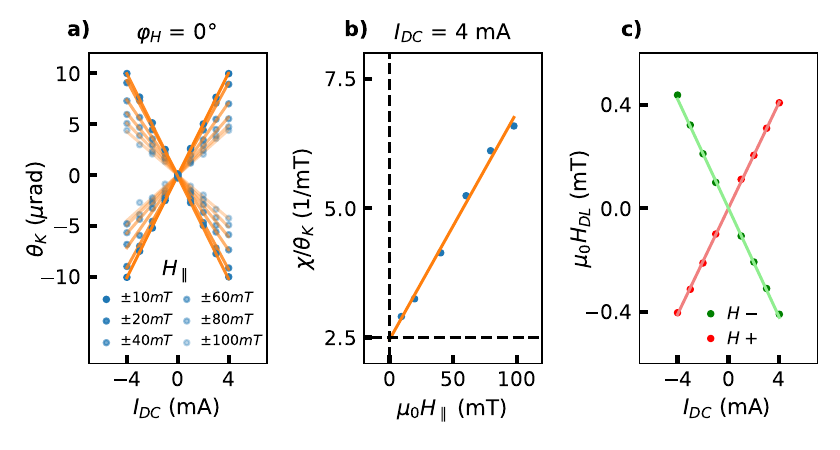}
\caption{
Determination of the effective magnetization and damping-like
spin-orbit torque in the Gd(2)/Pt(2)/Co(2)/Pt(2)
heterostructure, for which the approximation
$M_{\mathrm{eff}} \gg H_{\parallel}$ is not valid.
(a) Kerr rotation, $\theta_K$, as a function of dc current,
$I_{\mathrm{DC}}$, measured at $\phi_H=0^{\circ}$ for different
magnitudes of the applied in-plane magnetic field,
$\mu_0H_{\parallel}$. Solid orange lines are linear fits to the
data. (b) Linear dependence of $\chi/\theta_K$ on the applied
in-plane field, $\mu_0H_{\parallel}$, for
$I_{\mathrm{DC}}=4$~mA. The solid orange line is a fit to
Eq.~\ref{eq:Meff_fit}, from which both
$\mu_0H_{\mathrm{DL}}$ and $\mu_0M_{\mathrm{eff}}$ are
extracted. (c) Damping-like effective field,
$\mu_0H_{\mathrm{DL}}$, as a function of dc current for the two
opposite applied field directions, $\phi_H=0$, red curve and $\phi_H=\pi$, green curve. The
linear dependence of $\mu_0H_{\mathrm{DL}}$ on
$I_{\mathrm{DC}}$ is consistent with a current-induced
damping-like spin-orbit torque.
} \label{Fig:Sagnac3}
\end{figure}

To determine both $M_{\mathrm{eff}}$ and $H_{\mathrm{DL}}$,
we measure the Kerr angle as a function of dc current for several
values of the applied in-plane field, as shown in
Fig.~\ref{Fig:Sagnac3}(a). For a fixed current value,
Eq.~\ref{dl_eq} can be rewritten as
\begin{equation}
\frac{\chi}{\theta_K}
=
\frac{1}{\mu_0 H_{\mathrm{DL}}}
+
\frac{H_{\parallel}}
{\mu_0 M_{\mathrm{eff}} H_{\mathrm{DL}}},
\label{eq:Meff_fit}
\end{equation}
which predicts a linear dependence of $\chi/\theta_K$ on
$\mu_0 H_{\parallel}$. A linear fit to the data shown in
Fig.~\ref{Fig:Sagnac3}(b) therefore yields both
$\mu_0 H_{\mathrm{DL}}$ and $\mu_0 M_{\mathrm{eff}}$.

Among the samples studied here, only the
Gd(2)/Pt(2)/Co(2)/Pt(2) heterostructure required this more
general analysis because its effective magnetization is sufficiently small that $M_{\mathrm{eff}}$ is comparable to the applied in-plane field. Repeating the procedure for each current value yields $\mu_0 H_{\mathrm{DL}}$ as a function of
$I_{\mathrm{DC}}$ (Fig.~\ref{Fig:Sagnac3}(c)) and gives
$\mu_0 M_{\mathrm{eff}} = 56$~mT. Consequently, the approximation
$M_{\mathrm{eff}} \gg H_{\parallel}$ is not valid for this sample, and the full expression, Eq.~\ref{dl_eq}, was used.

\subsection{Transmission Electron microscopy in SFiM$_\mathrm{Pt}$}
High-resolution transmission electron microscopy (HR-TEM) was performed on a 10-$\mu$-wide Gd(2)/Pt(2)/Co(2)/Pt(2) strip to investigate the structural properties of the multilayer stack. The cross-sectional image reveals crystalline top Pt and Co layers, while the bottom Gd and Pt layers exhibit a polycrystalline structure. Partial intermixing between the Gd and Pt layers is observed at the interface, which may contribute to the enhanced resistivity of the multilayer compared to conventional Pt thin films.

\begin{figure}[h]
\centering
\includegraphics[width=0.47\textwidth]{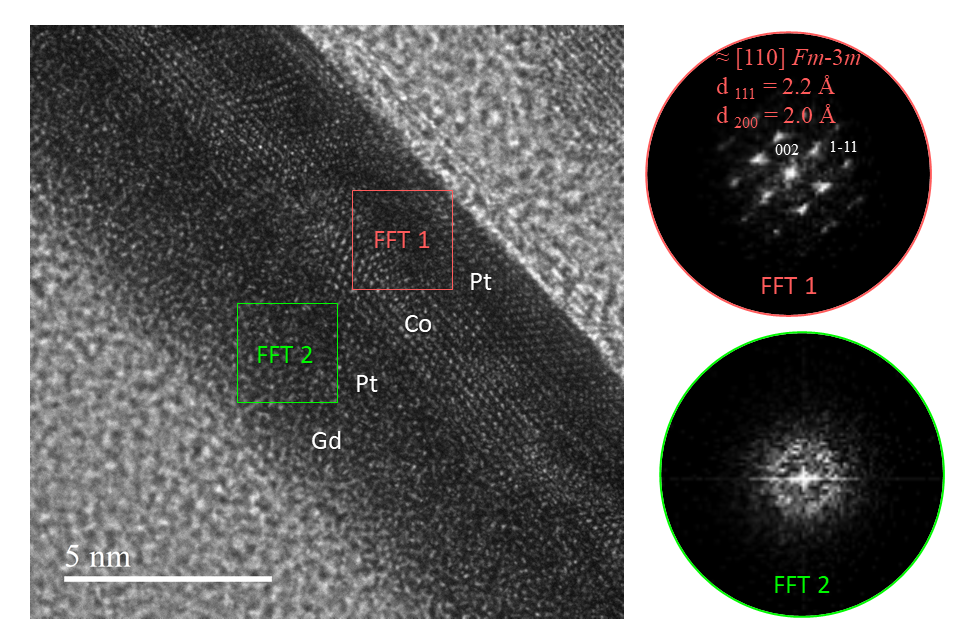}
\begin{quote}
\caption{Cross-sectional high-resolution TEM image of a patterned 10-$\mu$-wide Gd(2)/Pt(2)/Co(2)/Pt(2) device. The bottom layers (Gd and Pt) are disordered, as confirmed by the corresponding FFT (green rectangle), which shows a polycrystalline pattern, whereas the Co and top Pt layers exhibit crystalline order, as evidenced by their FFT (red rectangle))}\label{TEM}
\end{quote}
\end{figure}

\subsection{UV Lithography process in Gd/Pt/Co/Al}
\label{UV}
The harmonic Hall measurements were performed on six-terminal Hall bar devices with a channel width of 20~$\mu$m, whereas the Sagnac measurements were carried out on 20-$\mu$m-wide microstrips. Although fabricated from the same wafer, these two device geometries exhibit different magnetic anisotropy states. The microstrips preserve an anisotropy closer to that of the as-deposited films, with a dominant in-plane magnetic anisotropy, whereas the Hall bar devices exhibit an enhanced perpendicular magnetic anisotropy component. Consequently, the $\xi_{\mathrm{E}}^{\mathrm{DL}}$ values extracted from Sagnac and harmonic Hall measurements differ in magnitude, although the sign and current-induced torque symmetry remain consistent.
This anisotropy variation is attributed to differences in the UV lithography processing steps, particularly the exposure of the Co/Al interface to the developer solution. The microstrip fabrication involves two main steps: ion-beam etching followed by contact electrode deposition. In contrast, Hall bar fabrication includes an additional intermediate lithography step for SiO$_2$ deposition prior to the final electrode deposition. This SiO$_2$ layer serves as a protective insulating layer for subsequent gate-electrode fabrication. Because this step involves a lift-off process, the exposed surface of the current channel experiences a longer interaction with the developer, which acts as an Al etchant and modifies the metallic Al capping layer.
The resulting modification of the Co/Al/AlO$_x$ interface affects the interfacial magnetic anisotropy by changing the balance between metallic Al and oxidized Al at the interface. Previous studies have shown that the magnetic anisotropy of Co/Al-based systems is highly sensitive to the oxidation state and chemical environment of the interface~\cite{krishnia2025interfacial, manchon2008x}. Therefore, the additional lithographic exposure during Hall bar fabrication provides a mechanism for tuning the magnetic anisotropy toward the perpendicular direction.
Within the Hall bar devices, a systematic variation of the perpendicular saturation field is observed for channel widths of 4, 10, and 20~$\mu$m, as shown in Fig.~\ref{ALcapping}(a). This behavior can be understood considering the relative contribution of the exposed interface region. The developer penetration through the lift-off undercut is expected to be comparable for different device widths; however, this modified interface region represents a larger fraction of the total channel area in narrower devices. As a result, the 4-$\mu$m-wide Hall bars exhibit the strongest perpendicular anisotropy and enable the observation of current-induced SOT switching.
This sensitivity has practical consequences. Within the range of widths studied (4--20~$\mu$m), narrower devices are favorable for perpendicular anisotropy, but devices of different widths on the same chip are not magnetically equivalent, and the switching data in Fig.~\ref{ALcapping}(b) were obtained on devices from a single wafer; reproducibility across lithographic batches has not been tested. Possible routes to improve it are developers that do not etch Al, protection of the Al surface during lift-off (e.g., with a Ta layer), a modified order of the lithographic steps, and controlled oxidation or oxidation-resistant capping layers.

\subsection{ Field-sweep harmonic Hall measurements in a 4-$\mu$m-wide Gd/Pt/Co/Al Hall bar}
\label{FieldSweep} 

\begin{figure}[t]
\centering
\includegraphics[width=0.45\textwidth]{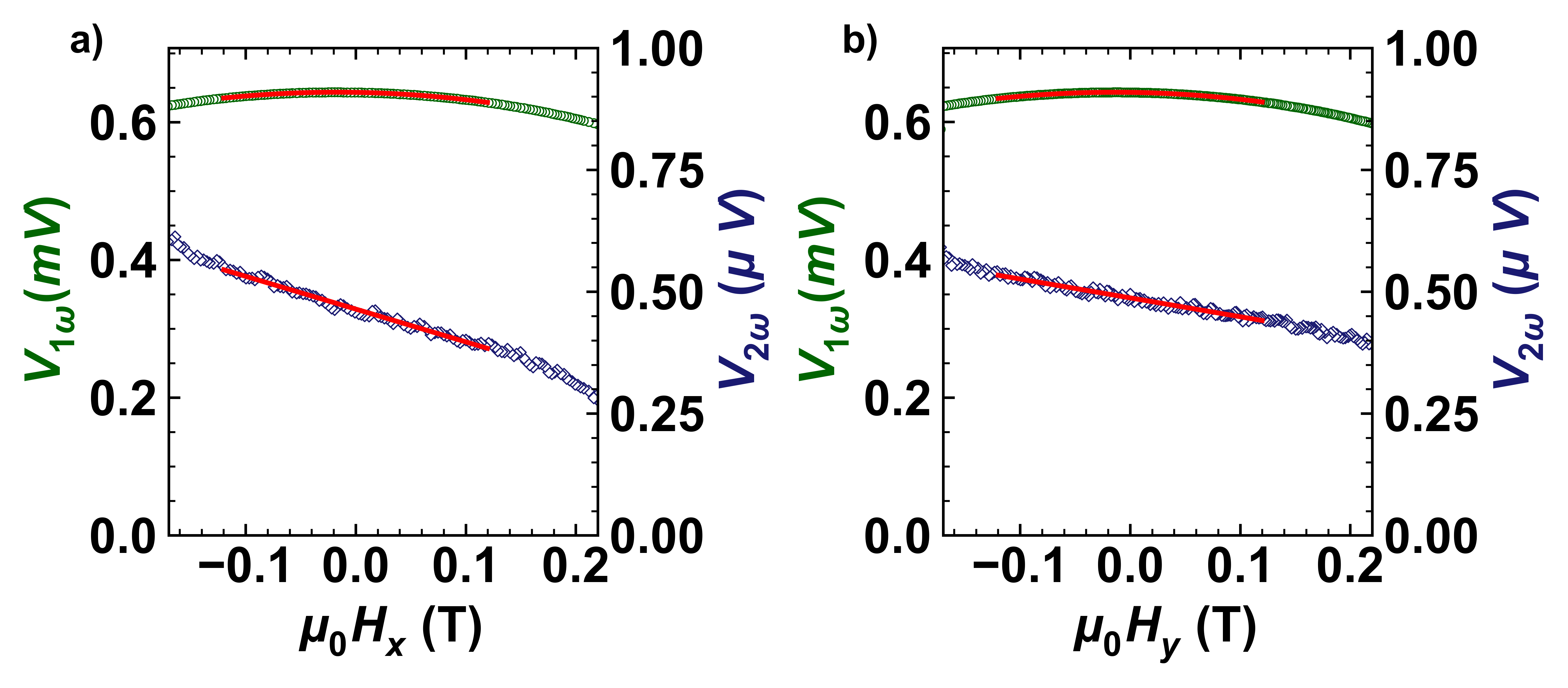}
\includegraphics[width=0.2\textwidth]{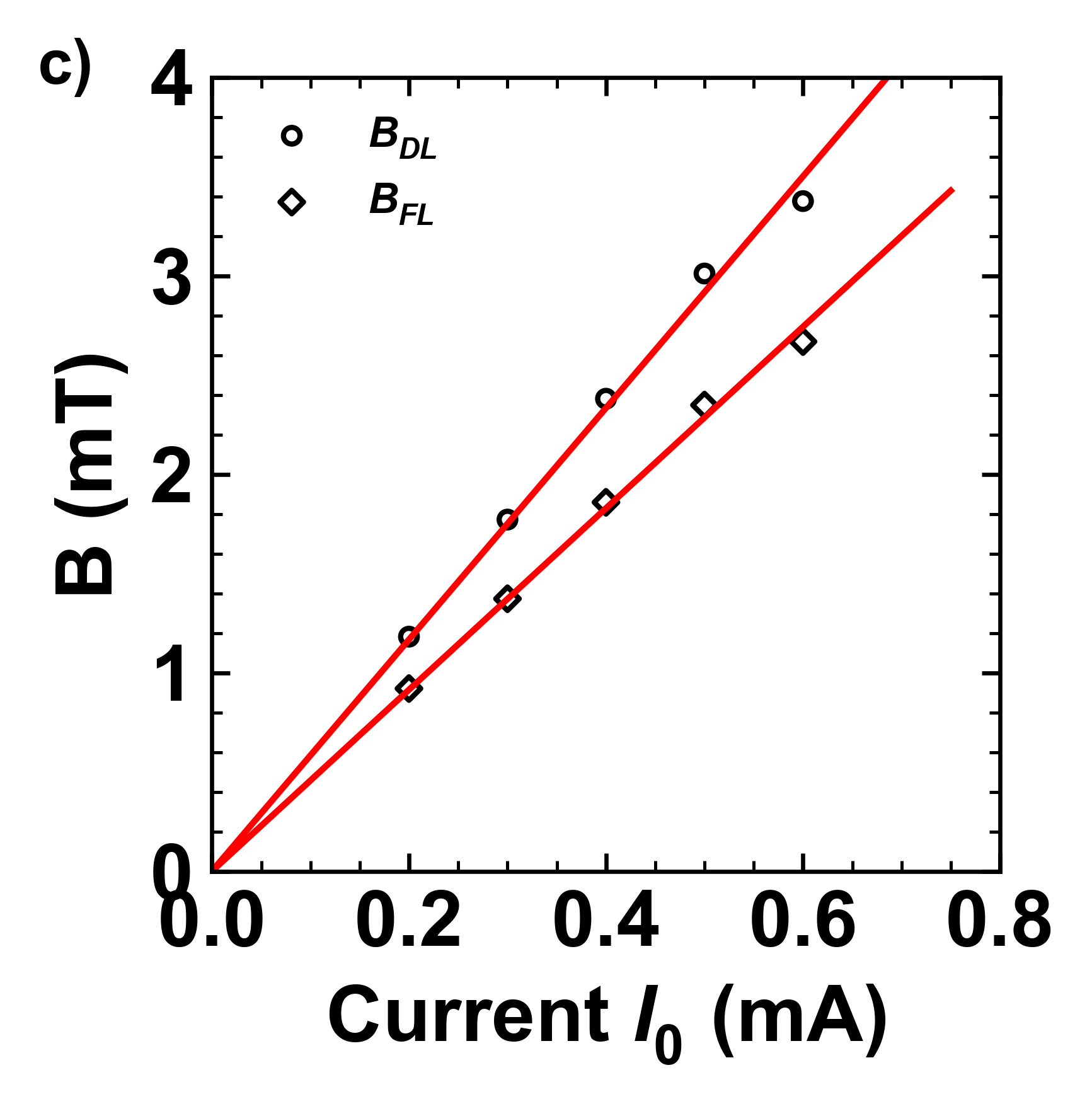}
\caption{Low-field harmonic Hall determination of the damping-like SOT in a 4-$\mu$m-wide Gd/Pt/Co/Al Hall bar with perpendicular magnetization. First-harmonic ($V_{1\omega}$, green, left axis) and second-harmonic ($V_{2\omega}$, blue, right axis) transverse voltages measured while sweeping the in-plane field (a) parallel to the current, $\mu_0H_x$, and (b) perpendicular to the current, $\mu_0H_y$, at a fixed current $I_0 = 0.2$~mA and frequency $433$~Hz. Solid red lines are fits to the data in the range $|\mu_0H| \leq 0.12$~T, used to extract the effective SOT field of Eqs.~(\ref{eq:dHx}) and (\ref{eq:dHy}). (c) Damping-like, $B_\mathrm{DL}$, and field-like, $B_\mathrm{FL}$,effective fields as a function of the applied current $I_0$. The red line is a linear fit through the origin, from which the field per unit current $H_\mathrm{DL}/I_0 = 5.83$~mT/mA is obtained. For clarity, only one magnetization state ($-M_z$) is shown.}
\label{fig:fieldsweep}
\end{figure}

As discussed above, the magnetic anisotropy of the Gd/Pt/Co/Al stack depends on the width of the Hall bar. For a width of 4~$\mu$m the magnetization is perpendicular to the sample plane (Fig.~\ref{ALcapping}a), and current-induced magnetization switching is also observed (Fig.~\ref{ALcapping}b). Since $\mathbf{M}$ points out of plane, the SOT efficiency can be determined with the low-field harmonic Hall protocol~\cite{hayashi2014quantitative,garello2013symmetry}, in which the first- and second-harmonic transverse voltages are measured while sweeping a small in-plane field either parallel to the current ($H_x$) or perpendicular to it ($H_y$). Using the independently measured anomalous and planar Hall resistances, $R_\mathrm{AHE}$ and $R_\mathrm{PHE}$, with $\xi = R_\mathrm{PHE}/R_\mathrm{AHE}$, the effective SOT field are given by
\begin{align}
B_{DL} &= 2\,\frac{\mathcal{H}_x + 2s\xi\,\mathcal{H}_y}{1-4\xi^2}, \label{eq:dHx}\\
B_{FL} &= 2\,\frac{\mathcal{H}_y + 2s\xi\,\mathcal{H}_x}{1-4\xi^2}, \label{eq:dHy}
\end{align}
where $\mathcal{H}_{x(y)} = \partial V^{2\omega}/\partial H_{x(y)} \big/ \partial^2 V^{\omega}/\partial H_{x(y)}^2$ and $s=\pm1$ denotes the direction of the out-of-plane magnetization. The field-like field $B_\mathrm{FL}$ includes the Oersted field. The quantities $\mathcal{H}_{x(y)}$ are obtained as shown in Fig. ~\ref{fig:fieldsweep}a. Repeating the procedure for different currents, we obtain $B_\mathrm{DL}$ as a function of the applied current, from which we extract the damping-like field per unit current, $B_\mathrm{DL}/I$ (Fig.~\ref{fig:fieldsweep}b). We find $B_\mathrm{DL}/I = 5.83$~mT/mA, corresponding to $\xi_\mathrm{E}^\mathrm{DL} = 1.43\times10^{5}~(\Omega\mathrm{m})^{-1}$ (Eq.~\ref{eff_DL_E}). This value is 3 times larger than that obtained on the 20-$\mu$m-wide device (Table~I), but is consistent with the low switching current density measured in this device (Fig.~\ref{ALcapping}b).

\end{document}